%% file: main.tex
\documentclass[acmsmall, screen]{acmart}

\AtBeginDocument{%
  }

\usepackage{multirow}
\usepackage{textcomp}
\usepackage{stfloats}
\usepackage{url}
\usepackage{verbatim}
\usepackage{graphicx}
\usepackage{xcolor}

\newcommand{\eg}{\textit{e}.\textit{g}.} 
\newcommand{\etal}{\textit{et} \textit{al}.}

\setcopyright{acmlicensed}
\copyrightyear{2025}
\acmYear{2025}
\acmDOI{XXXXXXX.XXXXXXX}

\acmVolume{1}
\acmNumber{1}
\acmArticle{1}
\acmMonth{1}

\begin{document}

\title{Urban Computing in the Era of Large Language Models}


\author{Zhonghang Li}
\email{bjdwh.zzh@gmail.com}
\orcid{0000-0002-3977-1334}
\affiliation{%
  \institution{South China University of Technology}
  \city{Guangzhou}
  \country{China}
}

\author{Lianghao Xia}
\email{aka\_xia@foxmail.com}
\affiliation{%
  \institution{The University of Hong Kong}
  \city{Hong Kong SAR}
  \country{China}
}

\author{Xubin Ren}
\email{xubinrencs@gmail.com}
\affiliation{%
  \institution{The University of Hong Kong}
  \city{Hong Kong SAR}
  \country{China}
}

\author{Jiabin Tang}
\email{jiabintang77@gmail.com}
\affiliation{%
  \institution{The University of Hong Kong}
  \city{Hong Kong SAR}
  \country{China}
}

\author{Tianyi Chen}
\email{tychen.cs@gmail.com}
\affiliation{%
  \institution{City University of Hong Kong}
  \city{Hong Kong SAR}
  \country{China}
}
\author{Yong Xu}
\authornote{Corresponding authors. Relevant resources will be maintained at \url{https://github.com/HKUDS/Awesome-LLM4Urban-Papers}}
\email{yxu@scut.edu.cn}
\affiliation{%
  \institution{South China University of Technology}
  \city{Guangzhou}
  \country{China}
}

\author{Chao Huang}
\authornotemark[1]
\email{chaohuang75@gmail.com}
\affiliation{%
  \institution{The University of Hong Kong}
  \city{Hong Kong SAR}
  \country{China}
}

\renewcommand{\shortauthors}{Li et al.}

\begin{abstract}
Urban computing has emerged as a multidisciplinary field that harnesses data-driven technologies to address challenges and improve urban living. Traditional approaches, while beneficial, often face challenges with generalization, scalability, and contextual understanding. The advent of Large Language Models (LLMs) offers transformative potential in this domain. This survey explores the intersection of LLMs and urban computing, emphasizing the impact of LLMs in processing and analyzing urban data, enhancing decision-making, and fostering citizen engagement.
We provide a concise overview of the evolution and core technologies of LLMs. Additionally, we survey their applications across key urban domains, such as transportation, public safety, and environmental monitoring, summarizing essential tasks and prior works in various urban contexts, while highlighting LLMs' functional roles and implementation patterns. Building on this, we propose potential LLM-based solutions to address unresolved challenges. To facilitate in-depth research, we compile a list of available datasets and tools applicable to diverse urban scenarios. Finally, we discuss the limitations of current approaches and outline future directions for advancing LLMs in urban computing.
\end{abstract}






\maketitle

\input{UCLLM_intro}

\input{UCLLM_relate}
\input{UCLLM_solution}

\input{UCLLM_data}
\input{UCLLM_future}
\input{UCLLM_conclusion}





\bibliographystyle{ACM-Reference-Format}
\bibliography{UCLLM_refers}


\end{document}

%% file: UCLLM_intro.tex
\section{Introduction}
\label{sec:intro}

In an era of rapid urbanization, cities worldwide face unprecedented challenges that stem from increasing population densities, resource constraints, and infrastructural demands~\cite{zheng2014urban_ub1}. Urban computing emerges as a pivotal interdisciplinary field that harnesses the power of computing technologies to address these complex urban issues. By integrating data acquisition, analysis, and modeling, urban computing endeavors to improve the quality of life in cities, enhance operational efficiencies, and promote sustainable urban development~\cite{liu2020urban_ub2,hashem2023urban_ub3,li2023gptst}. It is effective across a range of practical urban scenarios. For instance, in transportation, it enables the optimization of traffic flow through intelligent traffic signal control and real-time routing suggestions, thereby reducing congestion and emissions~\cite{du2023safelight_trans1}. Environmental monitoring leverages sensors and data analytics to track air and water quality, informing policy decisions and public health initiatives~\cite{zhang2023graph_environment}.

Deep learning, known for its robust representation and relationship modeling, plays a key role in urban computing. Urban data, including traffic, safety, and environmental metrics, exhibits strong temporal and spatial correlations. To capture temporal dependencies, researchers frequently utilize models such as Recurrent Neural Networks (RNNs)~\cite{lv2018lcrnn_rnn}, Temporal Convolutional Networks (TCNs)~\cite{wu2020connecting_MTGNN}, and attention mechanisms~\cite{song2020STSGCN}. Graph Neural Networks (GNNs) are widely used in spatial correlation analysis to model interactions between regions, enabling information transmission through message passing~\cite{li2024opencity,li2024flashst}. For tasks that require strategic decisions based on various environmental conditions, such as traffic signal control, reinforcement learning is commonly employed to optimize the decision-making process. 
Additionally, Convolutional Neural Networks (CNNs) are used in image recognition for tasks like land-use classification from satellite imagery~\cite{castelluccio2015landuse_cnn} and infrastructure anomaly detection~\cite{lei2023mutual_att}.

Despite these advancements, urban computing faced several bottlenecks that limited its full potential. 
\textbf{i) Multimodal data processing capabilities.} One significant challenge was the heterogeneity and complexity of urban data. Urban environments generate vast amounts of data that are diverse in nature, including numerical sensor readings, geospatial data, textual information from social media, and unstructured data such as images and videos. Integrating and analyzing this multimodal data to extract actionable insights proved difficult with traditional deep learning models, which often specialized in processing a single data type. 
\textbf{ii) Generalization ability.} The generalization of deep learning models is limited by temporal and spatial distribution shifts in urban data. Urban environments are diverse and dynamic, and models trained on historical data often fail to adapt to new patterns, reducing their effectiveness over time and across locations. This limits the deployment of robust, reliable solutions in evolving urban settings.
\textbf{iii) Interpretability.} The interpretability of deep learning models remained a concern. Decision-makers in urban planning and policy needed transparent and explainable models to trust and act upon the recommendations provided. Traditional deep learning models often functioned as black boxes, offering limited insight into the reasoning behind their predictions. 
\textbf{iv) Automatic planning capability.} The lack of automatic planning in urban computing limits practical applications, as predictions still rely on expert intervention to formulate responses. This dependence on experts slows decision-making and hinders real-time responses, emphasizing the need for autonomous planning systems.

In recent years, AI has undergone a transformative shift with the advent of LLMs, driving significant advances in natural language processing (NLP) and understanding. Notable models like ChatGPT~\cite{ouyang2022instructiontuning} and LLaMA~\cite{touvron2023llama} demonstrate human-like comprehension. 
Their success hinges on several key factors including model architecture, computational hardware, large-scale high-quality data, and advanced training techniques. Specifically, Transformers~\cite{vaswani2017attention} are fundamental to the architecture of LLMs, with their attention mechanism effectively capturing word relationships and handling long-range dependencies in parallel. Significant advancements in both performance and efficiency have been pivotal to the development of language models. Advancements in computational hardware and data expansion have enabled the scalability of language models~\cite{zhao2024survey_LLMsurvey1,chang2024asurvey_LLMsurvey2}. Research has progressively increased model parameters and data scale, exemplified by GPT-2~\cite{radford2019gpt2} and GPT-3~\cite{brown2020gpt3}, demonstrating the scalability of LLMs. Coupled with approaches like Reinforcement Learning from Human Feedback (RLHF)~\cite{stiennon2020RLHF}, these developments have produced sophisticated LLMs with robust understanding and reasoning abilities, significantly advancing the AI field.

\begin{figure}[t]
    \centering
    \includegraphics[width=0.98\textwidth]{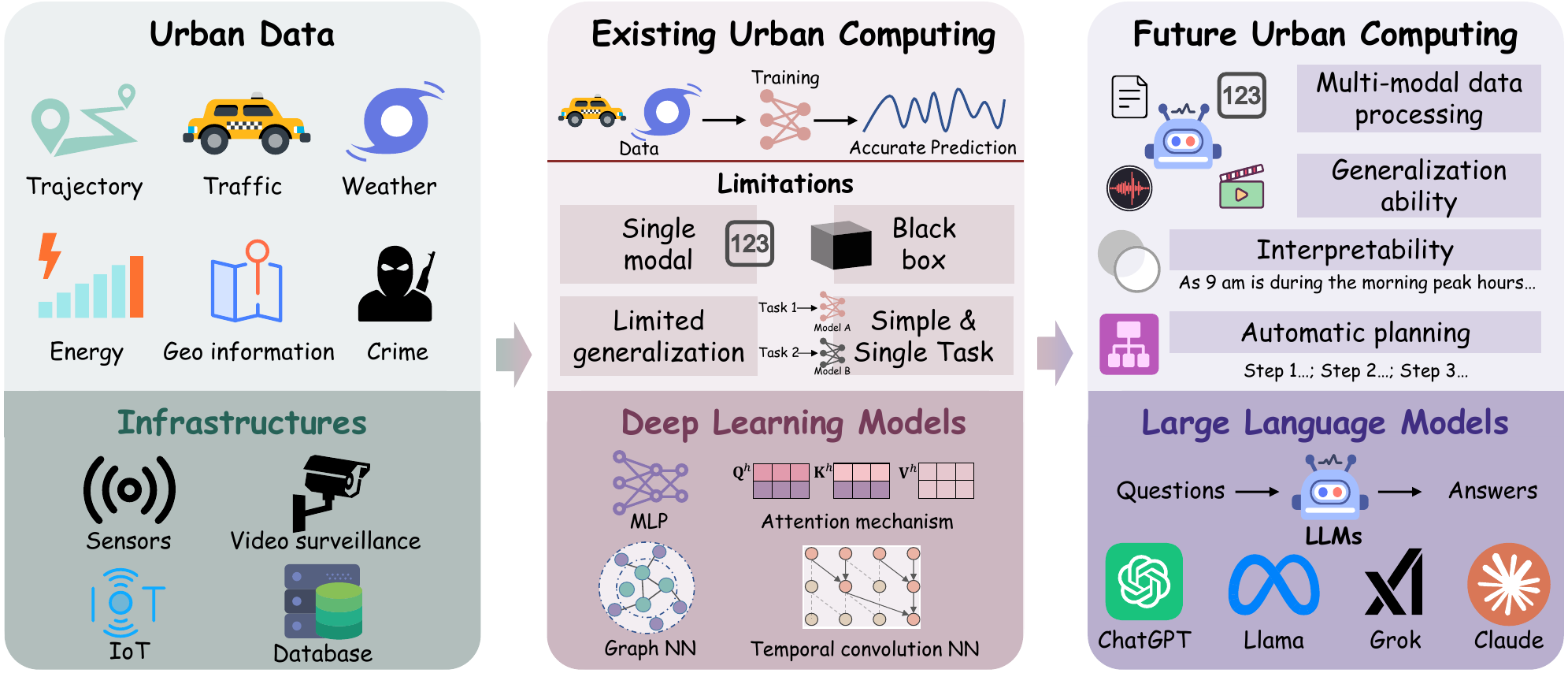}
    \caption{A sketch shows the evolution of urban computing technologies. The rapid advancement of Internet of Things (IoT) technologies and database enriched urban data (left). Despite deep learning techniques efficiently analyzing this data and producing useful insights, they struggle with multimodal data and autonomous planning (middle). The recent introduction of LLMs has advanced urban computing, highlighting their potential to improve urban efficiency (right).}
    \label{fig:intro1}
\end{figure}

LLMs have demonstrated utility beyond natural language processing (NLP), with applications in computer vision (CV)~\cite{liu2024llava}, software engineering~\cite{wadhwa2023softwareengineering}, and recommendation systems~\cite{ren2024rlmrec}. Recently, their potential roles in urban computing have also gained attention. For instance, in traffic prediction, LLMs excel by processing the textual temporal and geographical details in traffic data, thereby enhancing prediction capabilities~\cite{li2024URBANGPT}. They can also synthesize information from unstructured sources like reports and social media~\cite{han2024enhanced}, enabling rapid information extraction, trend identification, and sentiment analysis. Additionally, LLMs' reasoning and planning abilities have been explored in travel planning~\cite{chen2024travelagent}, streamlining processes and reducing human effort. With the swift advancements of LLMs in urban computing, a detailed and comprehensive review of pertinent studies is essential to showcase cutting-edge urban computing technologies and identify prevailing challenges. 
Fig.~\ref{fig:intro1} illustrates the evolution of urban computing technologies, highlighting the vital role of LLMs in shaping the future.

\emph{Related Surveys and Differences:} While numerous surveys have examined spatio-temporal modeling and prediction in urban computing, none have specifically focused on LLMs in this field. For example, survey~\cite{zou2025deep_survey1} discussed deep learning-based data fusion methods in urban computing until 2024, but did not specifically address LLM applications. \cite{jin2023large_survey2} reviewed large models for time series and spatio-temporal data up to 2023, whereas significant advancements in LLMs for urban computing emerged in 2023-2024. Additionally, some studies investigated foundation models in urban computing~\cite{zhang2024urban_surveyUF1,zhang2024urban_surveyUF2} and time series prediction~\cite{liang2024foundation_surveyTF}. As a crucial method in urban computing, spatio-temporal predictive techniques  have been extensive exploration. \cite{jin2024spatio_survey3} analyzed spatio-temporal GNNs for predictive learning, while other reviews have covered spatio-temporal GNNs applied to both general spatio-temporal~\cite{sahili2023spatio_surveyST1} and specific traffic data contexts~\cite{jiang2022graph_surveytraffic}. 
With the ongoing advancements in LLMs, their applications span diverse areas such as traffic management~\cite{zhang2024advancing_surveyTrans}, autonomous driving~\cite{yang2024llm4drive_surveyDrive}, and video understanding~\cite{tang2024videounderstanding_surveyvideo}. In this work, we provide a comprehensive analysis of LLM applications across nine urban computing subfields, highlighting their effectiveness and implementation in various urban contexts. Furthermore, we identify the challenges LLMs face in these sectors and propose prospective research directions and solutions.

\emph{Contributions:} The contributions of this survey are summarized as follows:
\begin{itemize}
\item To the best of our knowledge, this is the first survey that explores the applications of LLMs in various urban computing domains. It offers a comprehensive review of the literature and detailed discussions on how LLMs are applied across 9 urban scenarios, such as transportation and environmental monitoring, among others.
\item To ensure a thorough understanding, we have systematically reviewed task formats and previous approaches across various urban scenarios, as well as the development history and key technologies of LLMs. We have also categorized the roles that LLMs play in different urban domains-such as encoder, predictor, and agent-to comprehensively present the application scenarios and paradigms of LLMs across diverse domains.
\item We compile commonly used datasets from various domains in urban computing, thoroughly categorizing them and providing detailed information, including access links, data sources, and covered regions. Additionally, we summarize both traditional and emerging evaluation methods across these domains. This compilation significantly enhances experimental efficiency for researchers.
\item We summarize the challenges faced by LLMs in urban computing and identify future research directions. By analyzing the development status and bottlenecks of LLMs across various domains, we propose promising solutions and offer constructive recommendations for their future development in urban computing contexts.
\end{itemize}

\emph{Organization: }
This survey is structured as follows: Sec.~\ref{sec:relate} provides an overview of the development history and key technologies that have shaped LLMs. Sec.~\ref{sec:solution} explores the applications of LLMs within various urban computing contexts, divided into nine categories such as transportation, public safety, and environmental monitoring. Each category is discussed in a separate subsection, including: (i) definitions of tasks and an overview of prior research; (ii) an examination of how LLMs address specific domain challenges; and (iii) potential solutions proposed by our research to overcome these challenges. Sec.~\ref{sec:resources} reviews the datasets and evaluation methods commonly employed in different application scenarios. Sec.~\ref{sec:future} identifies the principal challenges faced by LLMs in urban computing and suggests directions for future research. The survey concludes with Sec.~\ref{sec:conclusion}.

%% file: UCLLM_relate.tex
\section{Background}
\label{sec:relate}

\subsection{Large Language Models (LLMs)}
LLMs mark a significant advancement in AI, demonstrating exceptional ability to understand and generate human-like text, and serving as a cornerstone for developing AI applications like automated writing assistants and advanced conversational agents. The introduction of the Transformer~\cite{vaswani2017attention} revolutionized NLP, outperforming RNNs in handling long-range dependencies. Since then, LLMs have evolved through key stages, each defined by unique traits and notable works.

The \textbf{Foundation Phase} laid the groundwork for modern LLMs, with BERT~\cite{devlin2019bert} introducing bidirectional context and pre-training techniques like Masked Language Modeling based on Transformer. Its success with 340M parameters highlighted the potential of large-scale language modeling, establishing the standard pre-training/fine-tuning paradigm. 
During the \textbf{Scaling Phase}, researchers focused on expanding model sizes and improving training methods. Representative works such as OpenAI's GPT-2 (1.5B parameters)~\cite{radford2019gpt2} demonstrated emergent capabilities through scaling, while GPT-3 (175B parameters)~\cite{brown2020gpt3} showcased advanced few-shot learning capabilities. In the \textbf{Capability Expansion Phase}, the advent of ChatGPT~\cite{ouyang2022instructiontuning} revolutionized NLP and AI by showcasing human-like conversation and task assistance. Its success spurred advancements in LLMs, leading to \textbf{\textit{commercial LLMs}} such as GPT-4(o)~\cite{openai2024gpt4technicalreport,gpt-4o}, o1~\cite{o1} and Claude~\cite{anthropic2024claude}, which demonstrate more astonishing situational understanding and reasoning abilities. 
In addition, the rise of \textbf{\textit{open-source LLMs}} has enabled researchers to explore and utilize LLMs across diverse domains. Advanced works, such as LLaMA~\cite{touvron2023llama}, Qianwen~\cite{yang2024qwen2} and Grok-1~\cite{XAI2024grok1}, have made significant contributions to the advancement of LLMs in various fields.

\subsection{Pioneering Technologies in LLMs Progress}
The success of LLMs across various domains can be attributed not only to increased training data and expanded model parameters but also to the following key technologies:

\noindent \textbf{i) Prompt Engineering}: This involves designing input text (prompts) to elicit desired outputs. Using prompt templates and few-shot examples, LLMs can accurately interpret user requirements and effectively perform specified tasks~\cite{white2023prompt,sahoo2024asystematic}.

\noindent \textbf{ii) Chain of Thought (COT)}~\cite{wei2022COT}: This method encourages models to "think out loud" by generating intermediate steps or reasoning paths before arriving at a final answer, enhancing the transparency and reliability of the outputs. 

\noindent \textbf{iii) Reinforcement Learning from Human Feedback (RLHF)}~\cite{stiennon2020RLHF}: RLHF leverages human feedback to refine model outputs by aligning them with high-quality standards. Feedback is used as training labels for reward models, which score LLM responses and guide them via reinforcement learning to produce superior outputs~\cite{bai2022training,kaufmann2024asurvey}.

\noindent \textbf{iv) Parameter-Efficient Fine-Tuning (PEFT)}: Techniques such as Adapter layers~\cite{houlsby2019Adapter} and Low-Rank Adaptation (LoRA)~\cite{hu2021LORA} allow for fine-tuning LLMs on specific tasks without altering all the parameters, thus saving computational resources and allowing for easier deployment.

\noindent \textbf{v) Instruction Tuning}~\cite{ouyang2022instructiontuning}: It improves model alignment with human intent by training on curated instruction-following datasets, enhancing task understanding and execution. It enables more reliable and controllable model behavior~\cite{zhang2024instructiontuningsurvey}.

\noindent \textbf{vi) Advanced Training Techniques}: Various advanced architectures have been proposed to enhance LLM performance and efficiency. For example, MoE~\cite{fedus2022switch,jiang2023mistral7b} uses specialized neural networks (experts) with a gating mechanism to route inputs, enabling efficient scaling and reduced computational costs. Flash Attention~\cite{dao2022flashattention} optimizes transformer attention by restructuring memory access patterns, reducing memory usage, and improving training speed. In addition, Rotary Position Embedding (RoPE)~\cite{su2024RoFormer} encodes positional information via rotation matrices, enabling better relative position modeling, unlimited sequence processing, and translation equivariance.

\noindent \textbf{vii) Retrieval-Augmented Generation (RAG)}~\cite{Lewis2020RAG}: It enhances response accuracy by combining LLMs with external knowledge retrieval. The model first retrieves relevant documents from a knowledge base, using them as context for generating responses. This approach reduces hallucinations and enables real-time knowledge updates without retraining. Futhermore, GraphRAG~\cite{edge2024graphrag} organizes knowledge into graph structures, leveraging entity relationships to improve context gathering and multi-hop reasoning. Subsequent works focused on enhanceing RAG's performance~\cite{kim2024autorag} and efficiency~\cite{guo2024lightrag}.

As LLM research advances, more powerful and versatile models are expected to emerge, enhancing the deployment of LLMs in samrt cities and improving urban management efficiency. Despite challenges, these developments significantly increase the potential of AI to benefit society.

%% file: UCLLM_solution.tex
\section{Urban computing with LLMs}
\label{sec:solution}
\begin{figure}[t]
    \centering
    \includegraphics[width=0.48\textwidth]{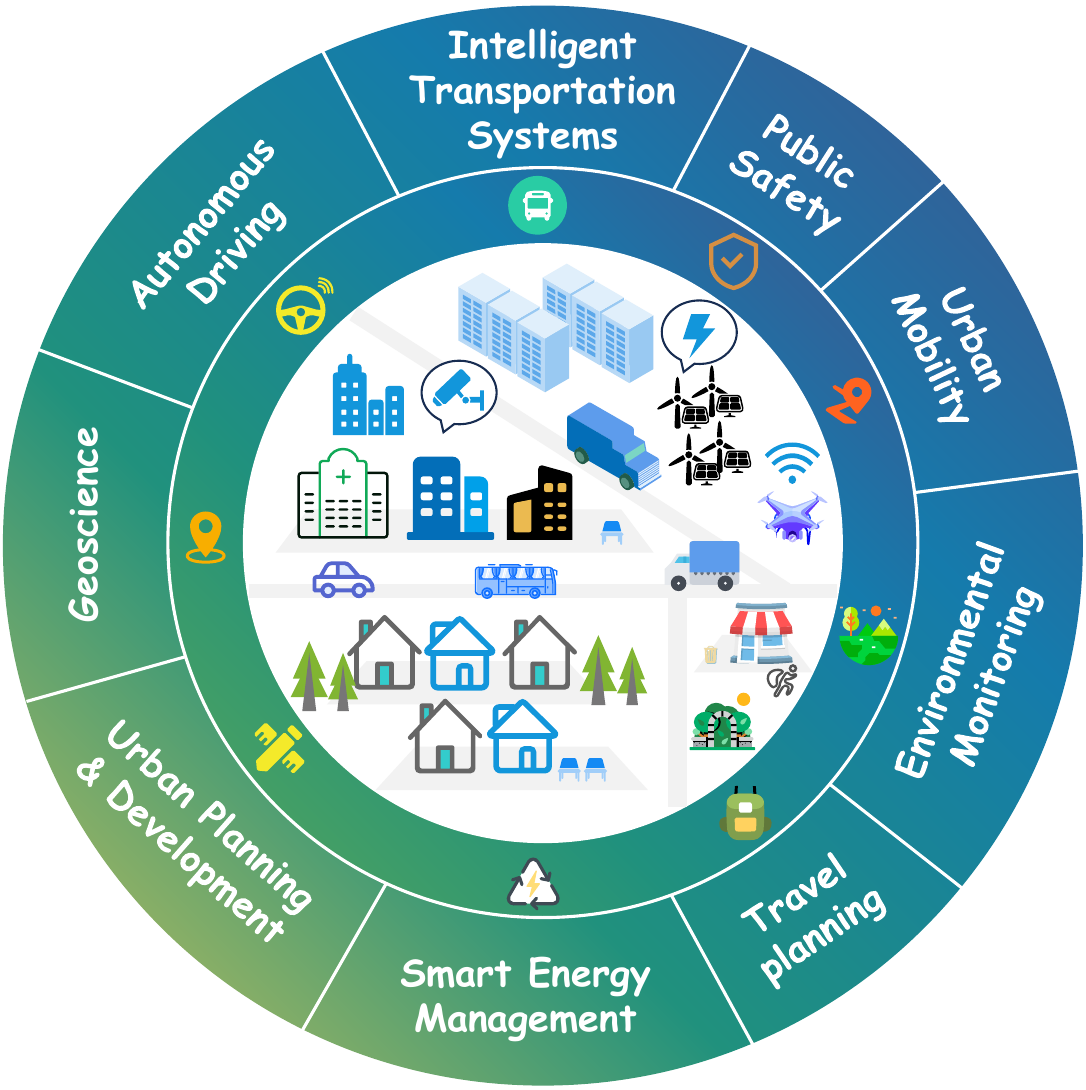}
    \caption{Various applications of LLMs within urban computing.}
    \label{fig:intro2}
\end{figure}

\begin{figure}[t]
    \centering
    \includegraphics[width=1\textwidth]{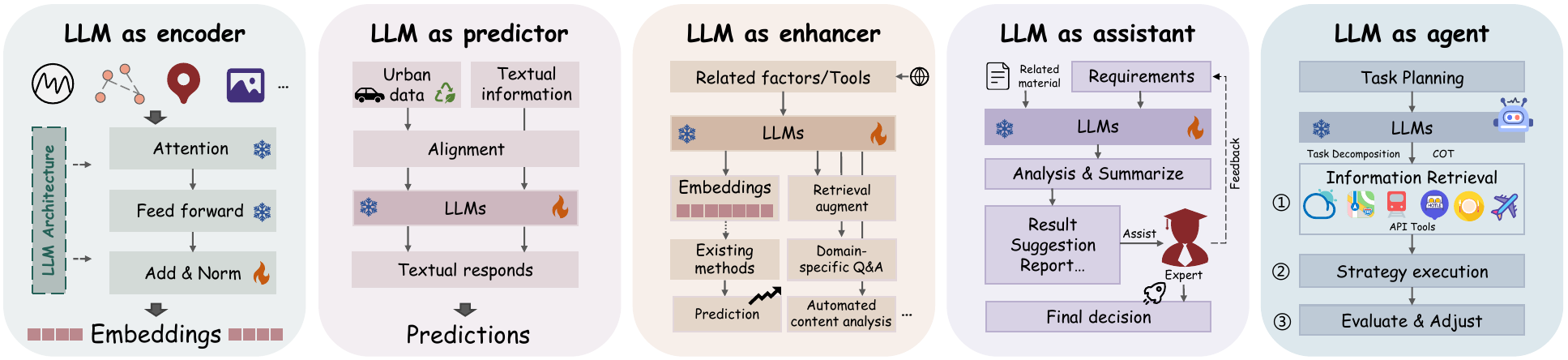}
    \caption{We categorize the roles of LLMs in urban computing into five functional types: encoder, predictor, enhancer, assistant, and agent.}
    \label{fig:llm_as}
\end{figure}

LLMs have shown remarkable capabilities in urban computing, providing innovative solutions and improving efficiency across domains. To clarify their applications, we classify tasks by application scenarios and explore how LLMs can be effectively utilized (Fig.~\ref{fig:intro2}). We begin by introducing task forms and reviewing prior work in urban computing scenarios. Next, we examine LLM applications, categorizing their roles - such as encoder and predictor (Fig.~\ref{fig:llm_as}) - while highlighting their impact on urban challenges. Building on prior research, we propose a blueprint for future LLM integration. Different solutions and their classifications are organized in Table~\ref{tab:categories}.

\subsection{Intelligent Transportation Systems}
Intelligent Transportation Systems (ITS) integrate information and communication technologies with transportation infrastructure to enhance traffic management and efficiency. By optimizing traffic lights and predicting traffic indicators, ITS reduces congestion and improves the sustainability and responsiveness of urban transportation networks.

\subsubsection{Traffic prediction} Traffic forecasting encompasses handling a variety of transportation tasks, including the prediction of traffic metrics (such as flow, speed, and indexes), traffic demand forecasting (like taxi and bike demand), and traffic imputation (filling in missing values).

\noindent \textbf{Core tasks \& previous works.} Traffic data can typically be represented by a three-dimensional tensor $\textbf{X} \in \mathbb{R}^{R \times T \times F}$. Here, $R$,$T$ and $F$ denote the number of regions, time steps, and features(\eg~inflow, outflow), respectively. A common form of traffic forecasting is predicting future traffic conditions over $P$ time steps based on the historical traffic states of $H$ time steps:
\begin{align}
    \label{eq:A1_formalized}
    \textbf{X}_{t_{K+1}: t_{K+P}} = g(\textbf{X}_{t_{K-H+1}: t_K})
\end{align}

Additionally, the traffic imputation task involves inferring missing traffic data based on partial observations, which can be formalized as follows: $\hat{\textbf{X}}=g(\textbf{X}, \textbf{M})$. Here, $\textbf{M} \in \{0, 1\}$ is the mask matrix and $\hat{\textbf{X}}$ denotes the reconstructed traffic state.
$g(\cdot)$ represents the traffic prediction methods which utilize spatio-temporal (ST) neural networks to model ST dependencies in traffic scenarios. For example, STGCN~\cite{yu2018spatio_STGCN} integrates gated temporal convolutional networks with GNNs, while GMAN~\cite{zheng2020GMAN} uses attention mechanisms and a learnable fusion layer to capture ST correlations.
Subsequent advancements in the field have focused on enhancing ST frameworks by incorporating real-world elements. This includes adopting multi-scale and multi-granularity temporal correlations in models such as MTGNN~\cite{wu2020connecting_MTGNN} and ASTGCN~\cite{guo2019attention_ASTGCN}, enhancing graph structures beyond predefined adjacency matrices in GWN~\cite{wu2019graph_GWN} and AGCRN~\cite{bai2020adaptive_AGCRN}, and developing delay-aware modules in PDFormer~\cite{Jiang2023PDFormer} to better simulate traffic flow dynamics.

\noindent \textbf{Enhanced Traffic prediction with LLMs.} The emergence of LLMs has revitalized traffic prediction by leveraging their robust capabilities and clear linguistic outputs. Current studies aim to improve model effectiveness, generalizability, and interpretability, grouped into three categories:

i) \emph{LLMs as encoder}: These models explore using LLM architectures as spatio-temporal encoders by employing data tokenization and fine-tuning to improve prediction accuracy. In traffic prediction, ST-LLM~\cite{liu2024spatial_STLLM} utilizes a partially frozen attention LLM as its backbone encoder, optimizing the model by fine-tuning parameters in layer normalization and attention layers. Similarly, models such as TPLLM~\cite{ren2024TPLLM}, STD-PLM~\cite{huang2024STDPLM}, and STTLM~\cite{ma2024spatial_STTLM} complete their model training with LoRA fine-tuning. Traffic data sometimes suffers from information loss due to issues like signal disruptions, leading to noise that affects predictions. To address this, GATGPT~\cite{chen2023GATGPT} integrates LLMs with a graph attention model for missing data imputation, while STLLM-DF~\cite{shao2024STLLMDF} uses a denoising diffusion probabilistic model for imputation before fine-tuning LLMs for precise future predictions.

ii) \emph{LLMs as enhancer \& encoder}: These models employ LLMs to encode spatio-temporal semantic information, thereby enhancing the predictive capabilities of downstream models. STG-LLM~\cite{liu2024how_STGLLM} tokenizes and concatenates textual descriptions with spatio-temporal information before feeding them into the LLMs, where fine-tuning is applied with frozen multi-head attention module parameters. In addition, STGCN-L~\cite{li2024spatio_STGCNL} enhances semantic capture by generating region-specific POI embeddings with the GPT-4 API. In urban delivery demands prediction scenarios, IMPEL~\cite{nie2024joint_IMPEL} utilizes LLMs to generate node representations and functional graphs based on geographic information texts, effectively compensating for the lack of historical data in new areas, thereby enhancing the model's zero-shot transferability. Additionally, GT-TDI~\cite{zhang2024semantic_GT-TDI} improves spatio-temporal frameworks by leveraging language models to create semantically informed tensors, which help in more effectively imputing missing values. 

iii) \emph{LLMs as predictor}: 
These models explore the capability of using LLMs as spatio-temporal predictors. UrbanGPT~\cite{li2024URBANGPT} addresses traffic prediction in data-scarce environments by integrating a spatio-temporal encoder and enriching prompts with context like times, city specifics, and POIs to enhance traffic pattern understanding. Predictive tokens are decoded via a regression layer and optimized through instruction-tuning for numerical predictions. xTP-LLM~\cite{guo2024towards_XTPLLM} introduces an interpretable framework using system prompts with traffic knowledge and COT reasoning. Numerical predictions are derived through text-based Q\&A with LoRA fine-tuning, while few-shot learning resolves discrepancies between predictions and interpretations during inference.

\subsubsection{Traffic management} Traffic management regulates vehicle and pedestrian flows to ensure safe and efficient transportation, using technology, planning, and infrastructure to reduce congestion, accidents, and travel times in urban areas.

\noindent \textbf{Core tasks \& previous works.} The primary goal of traffic management is to enhance Traffic Signal Control (TSC) by creating adaptive signaling strategies for varying traffic flows across lanes and intersections, thus improving traffic efficiency. TSC problem can be represented as a Markov Decision Process~\cite{pang2024illmtsc}, which can be formalized as follows:
\begin{align}
\label{eq:A2_formalized}
V^\pi(s) = \mathbb{E} \left[ \sum_{t=0}^\infty \gamma^t R(s_t, a_t, s_{t+1}) \mid \pi, s_0 = s \right]
\end{align}
where \( V^\pi(s) \) represents the value function under policy \( \pi \) and \( \gamma \) is the discount factor. \( R(s_t, a_t, s_{t+1}) \) is the reward received after transitioning from state \( s_t \) to state \( s_{t+1} \) due to action \( a_t \), and \( \mathbb{E} \) denotes the expected value considering all possible probabilities of state transitions and rewards. The optimization objective is to find an optimal policy \( \pi^* \) that maximizes the value function for all states \( s \), which is formulated as: $\pi^* = \arg \max_\pi V^\pi(s)$. Our goal is to develop a policy that maximizes rewards, typically consisting of minimizing average travel time, queue length, and waiting time. 

Current TSC methods are primarily divided into rule-based and reinforcement learning-based (RL-based) approaches. Rule-based methods utilize fixed principles to optimize signal strategies, such as implementing cyclic patterns~\cite{koonce2008traffic} or using greedy strategies to enhance road throughput~\cite{varaiya2013max}. On the other hand, RL-based algorithms employ objectives like average waiting time and queue length as reward functions, training agents to maximize these rewards through continuous interaction with the environment during training sessions~\cite{chen2020toward, pang2024scalable, wang2024unitsa}. 
RL-based methods have become the preferred approach for TSC due to their adaptability to dynamic traffic flows and effective traffic light timing allocation in recent years.

\noindent \textbf{Optimizing Traffic Management with LLMs.} 
The current application of large language models in traffic management primarily focuses on utilizing their advanced reasoning and planning capabilities to improve traffic management efficiency. This can be broadly classified into three categories:

i) \emph{LLMs as enhancer}: These works explore the potential of using LLMs to improve traffic management tasks. PromptGAT~\cite{da2024prompt_PromptGAT} uses simulation data, such as environmental traffic scene details, as input for an LLM, which generates dynamic traffic indicators reflecting real conditions. These indicators inform a reinforcement learning model to devise traffic management strategies for complex weather scenarios. iLLM-TSC~\cite{pang2024illmtsc} uses LLMs to evaluate and refine decisions made by RL agents in traffic scenarios, providing suggestions that are then applied in the simulation environment to update its state.
In addition, TransGPT~\cite{wang2024transgpt} has effectively addressed various traffic tasks such as traffic sign recognition, scene recognition, and driving advice by building a specialized model from scratch, using a large, multimodal traffic dataset for training.
LLMlight~\cite{lai2024llmlight} introduces LightGPT, a specialized model for TSC scenarios, fine-tuned with a critic network to effectively manage traffic light reasoning and decision-making within its framework using constructed prompts.

ii) \emph{LLMs as agent}: These models position LLMs as the central scheduler, incorporating external tools to enhance traffic management capabilities.
OpenTI~\cite{da2023openti} and TrafficGPT~\cite{zhang2024TrafficGPT} integrate external tools using prompt engineering and COT within their frameworks to efficiently plan and manage diverse traffic tasks such as geographic queries, navigation, and report analysis based on user requirements.
TP-GPT~\cite{wang2024traffic_TPGPT} uses a multi-agent collaborative approach for precise data retrieval and traffic report generation, while LA-Light~\cite{wang2024llm_LAlight} employs a specialized agent framework for TSC, leveraging perception and decision-making tools to gather real-time traffic data and suggestions for effective traffic signal control strategies.

iii) \emph{LLMs as assistant}: In these approaches, LLMs directly assist in completing traffic management tasks more efficiently. \cite{villarreal2023can} confirms that the LLM can effectively aid non-experts in traffic management by conducting a control experiment. \cite{tang2024large,dai2024large,masri2024leveraging} investigate how effectively human engineers and LLMs can collaborate to enhance traffic management, with most experiments showing that LLMs understand traffic scenarios and provide sensible suggestions.

\subsubsection{Future-Proofing Intelligent Transportation Systems} Based on existing research, we expect LLMs to enhance future intelligent transportation systems (ITS) in two main areas:
i) Enhancing data analysis: LLMs are adept at processing unstructured textual data from sources like social media, incident reports, and driver feedback, allowing ITS to quickly identify traffic incidents, road closures, or hazards from these inputs, improving response times and safety.
Improving Predictive Analytics for City Events: LLMs' ability to understand context and nuance in data makes them well-suited for predicting variations in traffic flow due to city events such as concerts, sports games, or sudden weather changes. By integrating LLMs with existing models, ITS can enhance traffic forecasts by incorporating model outputs and urban events. 
This supports proactive traffic management, such as dynamic signal adjustments and early route suggestions to ease congestion.

\subsection{Public Safety}
Urban computing in public safety utilizes advanced analytics to predict and mitigate urban risks like crimes and traffic accidents. It leverages real-time data to identify hazards, efficiently allocate resources, and deploy emergency services quickly. This enhances safety and sustainability in cities.

\noindent \textbf{Core tasks \& previous works.} As primary tasks in public safety, both traffic accident prediction and crime prediction are forms of spatio-temporal prediction, which can be represented by Eq~\ref{eq:A1_formalized}. Here, the feature dimension $F$ in the data $\textbf{X}$ typically includes categories of traffic accidents (\eg~minor or severe accidents) or types of crimes (\eg~robbery, theft). Consequently, they can also be structured as classification tasks focused on predicting the occurrence of specific events.

The traffic accident prediction task is usually enhanced by incorporating diverse data types such as satellite imagery, environmental factors, and statistics on risky driving behaviors for more comprehensive analysis. Studies~\cite{chen2016learning,najjar2017combining,yuan2018hetero,trirat2023MGTAR} utilize deep techniques like LSTM, CNNs, and GCNs for associative modeling across multiple factors.
In crime prediction, DeepCrime~\cite{huang2018DeepCrime} utilizes RNNs and attention mechanisms to capture dynamic patterns, while STSHN~\cite{xia2021spatial_STSHN} enhances attention networks to learn cross-category relationships. To address sparse data, STHSL~\cite{li2022spatial_STHSL} applies a ST self-supervised framework to model crime patterns locally and globally.

\noindent \textbf{Harnessing LLMs for Safer Cities} 
The exceptional associative and reasoning capabilities of LLMs make them ideal for analyzing complex public safety incidents, with their applications in this field falling into several key categories.

i) \emph{LLMs as encoder \& predictor.} Researchers have investigated the efficacy of leveraging LLMs as text encoders or for predictive tasks. 
\cite{grigorev2024enhancing} integrated LLMs with machine learning algorithms to enhance traffic accident severity classification, using LLMs to encode text for feature vector generation post-data preprocessing. \cite{sarzaeim2024experimental} tested BART and GPT's crime prediction in zero-shot, few-shot, and fine-tuned scenarios, finding GPT-4's zero-shot predictions superior to fine-tuned GPT-3 and random forest models in certain cases. UrbanGPT~\cite{li2024URBANGPT} leverages diverse urban data for instruction-tuning to learn city dynamics, enabling advanced zero-shot crime prediction.

ii) \emph{LLMs as assistant.} This line of research primarily investigates the capability of using LLMs as human assistants in public safety. 
In their design blueprint, \cite{zheng2023chatgpt} outlines LLMs' roles in traffic safety, suggesting automation of accident reports, traffic data enhancement, and safety analysis via sensors. WatchOverGPT~\cite{shahid2024WatchOverGPT} uses wearable devices and LLMs for real-time detection and response to criminal activities, centralizing LLMs in data processing and interaction.
Additionally, Several studies assess LLMs' and MLMs' reasoning abilities in public safety by using direct question inputs. \cite{mumtarin2023large} and \cite{zhen2024leveraging} analyze traffic accident data using LLMs to determine accident type, responsible parties, and severity. Similarly, \cite{hostetter2024the} explores LLMs’ applications in fire engineering using a text-based Q\&A method, while \cite{zhou2024gpt4v} inputs key incident frames into large visual language models to identify traffic events like accidents and violations.

iii) \emph{LLMs as enhancer.} Another segment of research has explored the application of LLMs in public safety with the enhancement of related tools. \cite{chen2024enhancing} developed a decision support system integrating LLMs with knowledge graphs, structuring emergency documents for better decision-making. \cite{Zarzà2023llm} combines deep networks for predicting traffic accidents with LLMs to provide behavior recommendations to drivers, while \cite{otal2024llmassisted} utilizs LLMs in emergency response systems to improve 911 dispatch efficiency by fine-tuning on disaster-related datasets. Additionally, \cite{zheng2023trafficsafetygpt} develops TrafficSafetyGPT, a traffic safety model refined with LLAMA-7B, utilizing the Road Safety Manual and ChatGPT for instruction generation.

\noindent \textbf{Securing Tomorrow with LLMs.} LLMs have the potential to serve as advanced traffic accident predictors. While traditional deep learning methods consider factors like weather and congestion, they often lack interpretability due to their 'black-box' nature. In contrast, LLMs process information at a higher semantic level, enabling a nuanced analysis of auxiliary factors and improving traffic accident probability assessments. Fine-tuning LLMs with COT prompting facilitates systematic analysis, making decision-making transparent by generating results along with explanations.

\subsection{Urban Mobility}
Urban Mobility prediction and understanding is a key aspect of urban computing, focusing on modeling and forecasting individual movements within cities based on historical patterns and factors such as time, urban structure, and activity purposes. This understanding is essential for various real-world applications such as traffic management and location-based services.

\noindent \textbf{Core tasks \& previous works.}
Urban mobility tasks aims to model, predict, and understand human movement patterns in urban environments. The core representation is a trajectory sequence $T = {(l_i,t_i)}^n_{i=1}$, where each point contains location $l_i$ and timestamp $t_i$. The primary task can be unified as a conditional generation problem: given historical observations $H$, contextual information $C$ (\eg~current stays, user profiles), and external knowledge $K$ (\eg~POI information, road networks), predict the next location(s) through a mapping function $f(H,C,K) \rightarrow {(l_{k+1},t_{k+1}),...,(l_n,t_n)}$. When $n=k+1$, this reduces to next location prediction~\cite{cheng2013you}; when $n>k+1$, it becomes trajectory generation~\cite{li2024more}. 
Additionally, the model can provide classification results for POIs or semantic interpretations of user activaties~\cite{luo2024deciphering, zhang2024large}. 

Traditional approaches to urban mobility prediction can be broadly categorized into several research lines. The most fundamental methods are statistical methods like Markov chains (\textit{e.g.}, FPMC~\cite{rendle2010factorizing}), which model mobility as transition probabilities between locations. With the rise of deep learning, sequence modeling approaches became prominent, represented by RNN-based methods such as DeepMove~\cite{feng2018deepmove} and LSTPM~\cite{sun2020go}. Graph-based methods emerged to better capture spatial relationships, with models like STP-UDGAT~\cite{lim2020stp}, HMT-GRN~\cite{lim2022hierarchical}, and DRGN~\cite{wang2022learning}. Some works also explored generative modeling perspectives, using VAEs for diverse trajectory generation and GANs (\textit{e.g.}, Social GAN~\cite{gupta2018social}) for synthesizing trajectories with social attributes. Additionally, mechanistic models like the gravity model~\cite{lenormand2016systematic} and radiation model~\cite{ren2014predicting} capture mobility patterns through explicit formulations considering population density and distance.

\noindent \textbf{LLMs Elevate Urban Mobility Insights.} The integration of LLMs has enhanced urban mobility prediction by leveraging their semantic understanding, reasoning capabilities, and extensive urban knowledge. Research in this area focuses on improving prediction accuracy, cross-city generalization, and model interpretability. We categorize these into three aspects:

i) \emph{LLMs as encoder \& enhancer}.
The first research direction leverages LLMs' strong semantic understanding capabilities to encode diverse urban mobility data into meaningful representations, divided into three sub-directions based on encoding targets: (1) Trajectory encoding, where papers like PLMTraj~\cite{zhou2024plm4traj} and ~\cite{liu2024semantic} utilize LLMs to extract semantic features from raw trajectory data and understand movement patterns; (2) POI encoding, demonstrated by works such as POI GPT~\cite{kim2024poi}, M3PT~\cite{yang2023m3pt}, and LARR~\cite{wan2024larr} that focus on converting POI-related information (\eg~textual descriptions, images, scene context) into rich semantic embeddings; and (3) Multi-modal encoding, represented by papers like ReFound~\cite{xiao2024refound} and CityGPT~\cite{feng2024citygpt} that combine multiple data modalities (\eg~text, images, spatial information) through LLM-based encoding. These approaches typically employ techniques such as contrastive learning and knowledge distillation to enhance the quality of encoded representations. The encoded outputs are then used for downstream tasks like POI classification, trajectory understanding, and recommendation. 
This encoding-focused approach captures semantic relationships and contextual information more effectively than traditional numerical encodings while preserving interpretability through LLMs' language understanding capabilities.

ii) \emph{LLMs as predictor}.
Recent research leverages LLMs as predictive models for urban mobility tasks using prompt engineering and few/zero-shot learning. For next location prediction, studies~\cite{wang2023would, feng2024move, beneduce2024large} demonstrate LLMs' ability to predict individuals' next visits without city-specific training. In aggregated flow prediction, works like LLM-COD~\cite{yu2024harnessing} and LLM-MPE~\cite{liang2024exploring} focus on collective mobility patterns and event-induced flows. POI recommendation, represented by LAMP~\cite{balsebre2024lamp} and LLM4POI~\cite{li2024large}, utilizes LLMs' semantic understanding for location suggestions.
These approaches typically employ carefully crafted prompting strategies and incorporate spatial-temporal constraints, with LLMs as predictors enhancing the ability to provide natural language explanations and manage complex contexts. They also show strong generalization across various cities and scenarios, overcoming a major limitation of traditional mobility prediction models.

iii) \emph{LLMs as agent}.
The third research direction treats LLMs as autonomous agents that understand and generate human mobility behaviors through advanced reasoning, adopting complex frameworks where LLMs perform multiple interactive roles beyond mere encoding or prediction. 
AgentMove~\cite{feng2024agentmove} uses decomposition-based methods to break complex mobility tasks into subtasks, systematically solved by LLMs.
Theory-guided agents, exemplified by CoPB~\cite{shao2024chain} and MobAgent~\cite{li2024more}, incorporate established behavioral theories (\eg~Theory of Planned Behavior) to guide LLMs in generating more realistic mobility patterns. Additionally, simulation-based approaches represented by LLMob~\cite{wang2024large} and MobilityGPT~\cite{haydari2024mobilitygpt} treat LLMs as virtual urban residents capable of exhibiting human-like mobility behaviors. 
A common theme in these approaches is the focus on semantic interpretability and real-world constraints. Agent-based frameworks excel in complex scenarios, such as during pandemics, and show enhanced generalization across urban environments, albeit with higher computational and design complexity.

\noindent\textbf{Future Work: From Prediction to Urban Intelligence}.
Future research will focus on hybrid collaborative frameworks combining the strengths of various approaches while addressing their limitations. Key areas include the development of multi-agent systems where specialized LLM agents manage different mobility modeling tasks-from semantic understanding to prediction and reasoning-linked through a standardized communication protocol and enhanced by RAG for accessing real-time urban data. Another area is causality-aware mobility frameworks that predict movements and elucidate the causal factors behind human mobility, integrating behavioral and psychological models. In addition, privacy-preserving modeling using federated learning and enhancing mobility understanding with multimodal inputs like satellite imagery and environmental sensors are also promising directions.

\subsection{Environmental Monitoring}
Smart environmental monitoring uses sensor networks to track city environmental conditions like air quality, disasters and weather. This data-driven approach enables real-time assessment of urban environments, helping authorities make informed decisions about public health and environmental management while supporting sustainable city development.

\noindent \textbf{Core tasks \& previous works.} 
Environmental monitoring encompasses various tasks, with the primary task of predicting future environmental conditions from historical data. This can be formalized as Eq~\ref{eq:A1_formalized}, where the feature dimension $F$ of $\textbf{X}$ represents the number of environmental factors, such as meteorological information, air quality indices (AQI) and water quality metrics.

Deep learning is increasingly vital in environmental monitoring, with applications ranging from air quality prediction to land cover classification, biodiversity assessment, and climate modeling~\cite{magazzino2024impact}. For example, CNNs have demonstrated high efficiency in using satellite imagery for detecting land use changes, mapping vegetation, and monitoring deforestation~\cite{yuan2020deep}. DeepAir~\cite{yi2018deep} proposes a deep learning method for air quality prediction by utilizing multiple types of data (\eg~meteorological, pollution data), while AirFormer~\cite{liang2023airformer} leverages Transformers to accurately predict air quality across China with detailed granularity, significantly reducing errors compared to existing models. These developments underscore deep learning's transformative role in environmental science, providing advanced solutions for sustainability.

\noindent \textbf{LLMs Aid in Environmental Monitoring.} The integration of LLMs in environmental monitoring has advanced significantly, using their natural language processing skills to improve data analysis, prediction, and decision-making. This section outlines four key roles of LLMs in this field as follows:

i) \emph{LLMs as predictor \& encoder.}
LLMs enhance environmental monitoring by effectively predicting variables such as wind speed and city temperature, utilizing their advanced language processing abilities. PromptCast~\cite{Xue2024PromptCast} and STELLM~\cite{wu2024STELLM} are innovative LLM applications in environmental monitoring. PromptCast transforms time series forecasting into sentence generation tasks by converting data into prompts, while STELLM improves wind speed forecasting by separating wind series into components (\eg~seasonal, trend) and integrating prompts to enhance accuracy.

ii) \emph{LLMs as assistant.} LLMs serve as virtual assistants in environmental monitoring, using their vast knowledge and reasoning skills to address complex environmental queries. For instance, \cite{xia2024question} creates a T5-based Q\&A service for typhoon disasters, enhancing retrieval accuracy and interaction with domain fine-tuning and RAG. \cite{sun2023unleashing} develops the ZFDDA model for visual question answering in flood damage assessment, using chain of thought demonstrations to improve disaster response without pre-training. QuakeBERT~\cite{han2024enhanced}, an LLM fine-tuned to classify social media texts, enhances earthquake impact analysis by assessing physical and social impacts, showcasing LLMs' ability to analyze unstructured data for disaster management. Recent developments in climate-focused LLMs have produced several specialized tools. Arabic Mini-ClimateGPT~\cite{mullappilly2023arabic} focuses on climate conversations in Arabic, combining fine-tuned responses with vector embedding retrieval for enhanced information access. In addition, ClimateBert~\cite{leippold2022climatebert}, pretrained on climate related texts, shows improved accuracy in tasks like classification and fact-checking. Similarly, climateGPT-2~\cite{vaghefi2022deep} enhances climate-related text generation by fine-tuning GPT-2 on climate change texts.

iii) \emph{LLMs as enhancer.} The capacity of LLMs to enrich environmental monitoring systems is evident in their application as enhancers. Researchers have integrated LLMs with Geographic Information Systems (GIS) to develop frameworks that improve public flood risk perception by enriching LLMs with flood knowledge~\cite{zhu2024flood}, showcasing their potential to enhance risk understanding through natural language dialogues. Besides, ChatClimate~\cite{vaghefi2023chatclimate} incorporates IPCC AR6 reports into large LLMs to provide reliable climate information for decision-making. 
Moreover, LLMs are integrated with knowledge-based systems to enhance adaptive fire safety planning, enriching expert systems with specialized reasoning for dynamic environments~\cite{durmus2024role}. 
In the context of disaster response, \cite{wang2024Near} introduces a framework for near-real-time fatality estimation following earthquakes. This framework proposes a LLM-based hierarchical model for casualty extraction from multilingual social media, with a dynamic truth discovery model for accurate and timely estimates, demonstrating LLMs' adaptability in complex disaster scenarios.

iv) \emph{LLMs as agent.} LLMs serve as autonomous agents that analyze extensive environmental data, providing actionable insights and improving decision-making. ~\cite{kraus2023enhancing} proposes an agent frmaework that accesses various data sources like ClimateWatch and internet searches to retrieve emission data and provide precise climate information. This prototype agent highlights the potential of LLMs to effectively integrate real-time information from various external tools, delivering more reliable and accurate environmental intelligence monitoring.

\noindent \textbf{Deeper integration of LLM agents with environmental monitoring.} The potential of LLM agents in environmental monitoring is significant, with anticipated advancements aimed at closing the gap between complex data analysis and actionable insights. As these agents evolve, they are expected to better integrate with various tools and platforms, enhancing their ability to process and analyze multimodal data streams for deeper environmental intelligence. Future multimodal LLM agents will autonomously access and analyze diverse data types—text, images, and sensor information—collaborating with other AI systems to deliver comprehensive monitoring solutions. This will support their role in early detection, predictive modeling, and real-time responses, leading to more effective environmental management.

\subsection{Travel planning}
Travel planning organizes trip essentials such as lodging and itineraries, reducing stress and optimizing budgets for smoother transitions and more efficient journeys. It improves quality of life by enabling cost-effective travel and maximizing experiences for leisure, work, or commuting.

\noindent \textbf{Core tasks \& previous works.} Travel planning can be formalized as a multi-objective optimization problem aimed at maximizing satisfaction $S$ (\eg~comfort levels, preference matching) and minimizing total cost $C$ (\eg~transportation, accommodation), subject to various constraints:
\begin{align}
\label{eq:A5_formalized}
\zeta = \min(\alpha \sum C + \beta \sum T) - \max(\gamma \sum S)
\end{align}
where $\alpha$, $\beta$, and $\gamma$ are weighting coefficients and $T$ represents time-related factors. The constraints include a budget limit $B$ such that $\sum C \leq B$; a time window $[t_s, t_e]$ where $t_s$ is the start time and $t_e$ is the end time; location constraints $L(x,y)$ defining geographical boundaries; and capacity constraints $K$ for accommodations and activities. Additionally, travel-related common sense and user preferences are part of the constraints.

Before the emergence of LLMs, researchers typically addressed this task as a POI planning or recommendation problems. To better tailor plans to users' travel scenarios, related studies framed it as an Orienteering Problem, considering factors such as starting and ending locations and times~\cite{taylor2018travel,chen2014automatic,zhang2018itinerary}. Additionally, POI recommendation systems utilized deep learning models trained on users' historical data to enhance next-POI suggestions~\cite{xu2024mmpoi,zheng2024decentralized}. Despite achieving some success, these systems offer only basic suggestions without generating complete travel itineraries.

\noindent  \textbf{Employing LLMs for Detailed Travel Planning.} The applications of large language models in travel planning can be categorized into four main areas:

i) \emph{LLMs as encoder.} This research area focuses on improving LLMs for journey planning using travel-specific datasets. \cite{meyer2024acomparison} develops a chatbot, fine-tuned on the Reddit travel dataset with QLoRA and Retrieval Augmented Fine-Tuning (RAFT) respectively, and further improves with Reinforcement Learning from Human Feedback, where RAFT demonstrats superior performance. TourLLM~\cite{wei2024tourllm} leverages Qwen model, applying LoRA fine-tuning on a diverse travel dataset to better introduce tourist attractions and assist in itinerary planning.

ii) \emph{LLMs as predictor.} This category utilizes the inferential capabilities of LLMs for travel-related tasks. For example, through experimentation, ~\cite{barandoni2024automating} found that GPT-4 and Mistral-7B were the top-performing models among commercial and open-source options, respectively, for extracting travel needs from TripAdvisor posts. 
\cite{zhai2024enhancing} and~\cite{mo2023large} demonstrated that incorporating travel-related information, such as user profiles, travel contexts, and historical data, into LLMs could improve the accuracy of transportation choice predictions. Although these LLMs were not trained on targeted datasets, their performance can sometimes surpass that of traditional methods.

iii) \emph{LLMs as agent.} These studies focus on enhancing travel planning by integrating LLMs with external tools and defining travel tasks textually to streamline journey planning. For instance, TravelPlanner~\cite{xie2024TravelPlanner} and TravelAgent~\cite{chen2024travelagent} enhance LLM capabilities by incorporating multi-dimensional constraints (\eg~personal hard constraints, soft constraints, common-sense constraints) and invoking external tools (\eg~map APIs, distance calculators) to plan optimal sightseeing routes and generate multi-day itineraries based on user conditions.

Another line of research focuses on integrating LLMs with planning algorithms to improve personalized travel planning. For instance, \cite{hao2024large} employs LLMs to convert travel requirements into SMT problems and generates Python code to operate SMT solvers for itinerary suggestions. Similarly, \cite{li2024research} utilizes LLMs to extract key data from travel needs and incorporate this into optimization equations, using genetic algorithms to optimize travel routes by considering time, distance, and satisfaction. ITINERA\cite{tang2024itinera} analyzes user requests to perform similarity searches in a POI database, clusters POIs geographically, and determines optimal travel sequences with a hierarchical Travelling Salesman Problem (TSP) approach. TRIP-PAL\cite{rosa2024trippal} combines LLMs with automated planning algorithms, using LLMs to gather travel information such as attractions and durations, which are then used to parameterize automated planners for creating optimal travel plans.

\noindent \textbf{The Next Era of Travel planning with LLMs.} LLMs in travel planning can be improved in several key areas:
i) Multimodal information utilization: Visual data, such as images and videos, are essential to travel planning, as they convey critical details about attractions (\eg~atmosphere), accommodations (\eg~room features), and restaurants (\eg~food presentation). Developing multimodal, travel-focused large models that can effectively handle such diverse data types could yield more comprehensive travel recommendations.
ii) Integration of booking links: Embedding direct booking links for transport, accommodations, and attractions into travel plans can increase user convenience. This could be enabled by multi-tool integration managed by agents.
iii) Edge-based tools: To handle sensitive personal data securely, future research should focus on using knowledge distillation and model compression techniques to develop edge-based LLMs for travel planning, which allow for private and efficient local processing.

\subsection{Urban Planning and Development}
In urban planning, it processes diverse data sources like traffic, population density, and infrastructure usage to optimize city design, improve resource allocation, and enhance quality of life. This data-driven approach enables evidence-based decision-making for sustainable urban development.

\noindent \textbf{Core tasks \& previous works.} Tasks in intelligent urban planning can generally be decomposed into distinct categories: classification tasks (\eg~region type identification), regression tasks (\eg~population density estimation), question-answer tasks (\eg~knowledge responses), and complex systematic optimization tasks (\eg~solution generation).

Deep learning has revolutionized urban planning by providing innovative solutions to complex challenges.~\cite{yao2019ahuman} used CNNs and random forests for street environmental analysis, while~\cite{sideris2019using} developed a machine learning method to assess building suitability.~\cite{yao2020delineating} introduced a deep bag-of-features network to analyze urban employment and housing patterns.~\cite{lv2014traffic} applied RNNs for traffic flow prediction, aiding in the allocation of public transportation resources. In addition,~\cite{yuan2022finegrained,wu2024deep} demonstrated that deep learning models leveraging historical and geospatial data can accurately predict land use changes, aiding urban planning. These advancements streamline processes and foster sustainable city development.

\noindent \textbf{LLMs Driving Urban Development.} The integration of LLMs in urban planning marks a significant advancement, leveraging natural language processing to enhance urban design, management, and sustainability. This section reviews these applications focusing on the following roles:

i) \emph{LLMs as assistant.}
LLMs enhance urban planning by leveraging vast knowledge and reasoning abilities to process large data volumes for insightful decisions. HSC-GPT~\cite{ran2023hsc}, trained on multidisciplinary datasets in urban planning, landscape gardening, and architecture, understands spatial semantics and generates innovative design solutions. UrbanLLM~\cite{jiang2024urbanllm}, fine-tuned on LLAMA-2-7b, autonomously handles urban planning by decomposing queries into subtasks, selecting AI models, and synthesizing responses, reducing reliance on human experts and improving efficiency.

ii) \emph{LLMs as agent.} LLMs are emerging as autonomous agents in urban planning. ~\cite{zhou2024large} proposes a multi-agent framework simulating participatory planning, where LLM agents represent planners and communities to generate land-use plans addressing community needs. Implemented in Beijing, it outperforms human experts in service accessibility and ecological metrics while improving resident satisfaction. PlanGPT~\cite{zhu2024plangpt}, a specialized LLM for urban planning, integrates database retrieval, domain-specific fine-tuning, and advanced tools to tackle planning challenges. Experiments show its proficiency in generating texts, retrieving information, and evaluating documents. These studies highlight LLMs' transformative role in participatory and efficient urban planning.

\noindent \textbf{Shaping urban planning with LLM innovations.} Future LLMs will revolutionize urban planning through two key capabilities. First, they can analyze large volumes of planning documents, impact reports, and public feedback, extracting insights often missed in manual reviews. Second, multimodal LLMs can process diverse data types, such as satellite imagery, CAD drawings, and GIS maps, enabling comprehensive analysis of spatial relationships and development patterns. This integration supports decisions in subway route optimization, land use allocation, and zoning regulations, while considering factors like population density, environmental impact, and economic viability.

\subsection{Smart Energy Management}
Smart Energy Management in urban computing uses IoT sensors, data analytics, and AI to enhance energy efficiency and sustainability in cities. This approach optimizes energy use, reduces consumption, and supports green initiatives, aiding in the development of energy-smart cities.

\noindent \textbf{Core tasks \& previous works.} The core of smart energy management is predicting energy usage, similar to general time series forecasting, which projects future data based on historical data and various feature variables, as described in Eq ~\ref{eq:A1_formalized}. Compared to \underline{S}patio-\underline{T}emporal forecasting, which accounts for both spatial and temporal correlations ($\textbf{X}_{ST} \in \mathbb{R}^{R \times T \times F}$), \underline{T}ime \underline{S}eries forecasting concentrates solely on temporal relationships ($\textbf{X}_{TS} \in \mathbb{R}^{T \times F}$). Energy forecasting is considered a principal task in time series prediction, with existing studies frequently utilizing datasets on power load (\eg~ETT-X) and electricity consumption (\eg~Electricity) for model evaluation.

Existing methods use temporal learning networks for energy consumption analysis. For instance, \cite{bacanin2023multivariate} merges RNNs with swarm intelligence for power grid forecasting, while \cite{nascimento2023a} combines a Transformer with a wavelet transform for wind energy prediction. Transformer's encoding capabilities have driven research in Transformer-based time series prediction, leveraging attention mechanisms to capture correlations in energy data~\cite{zhou2021informer,wu2021autoformer,zhou2022fedformer,wu2023timesnet}, advancing energy management.

\noindent \textbf{Enhancing Energy Management with LLMs}. LLMs are extensively used in time series forecasting, especially in energy prediction, serving various roles across different frameworks:

i) \emph{LLMs as encoder}. This research line employs LLMs as encoders to effectively extract temporal dependencies in energy data. For instance, \cite{zhou2023ofa} employs GPT-2 as the time series encoder, where it freezes the attention layers and feed-forward layers during training and employs a linear layer to decode the forecasted values. LLM4TS~\cite{chang2024llm4ts} applies tuning strategies like PEFT and full parameter fine-tuning to adapt LLMs for time series data. Additionally, The effectiveness of LLMs as encoders is assessed through ablation experiments in \cite{tan2024are}, which suggest that LLMs perform comparably to regular encoders in time series forecasting.

ii) \emph{LLMs as enhancer \& encoder}. Another line of research explores the use of LLMs' linguistic understanding to enhance the modeling of temporal dependencies. Autotimes~\cite{liu2024autotimes} segments energy data and uses frozen LLMs for positional encoding on these segments based on the temporal textual information, which, along with integrated embeddings, are input into the LLMs for decoding predictions. TEMPO~\cite{cao2024tempo} employs a prompt mechanism to direct the LLM's focus toward components of time series, such as trends, seasonality, and residuals, and fine-tunes the model using LoRA to extract hidden representations of these components. In addition, UniTime~\cite{liu2024unitime} uses language models to identify domain characteristics, reducing multi-domain confusion in predictions. TIME-FFM~\cite{liu2024timeffm} combines federated learning with text-modal alignment techniques for domain knowledge sharing and privacy-preserving time series prediction.

Due to the representational gap between time series/energy data and textual descriptions, bridging these two data types is crucial. TIME-LLM~\cite{jin2024timellm} utilizes patch reprogramming to convert time series block embeddings into text prototypes understandable by LLMs, employing cross-attention mechanism to align these embeddings. These embeddings are then concatenated with prompts for input into a frozen LLM. TEST~\cite{sun2024test} aligns time series embeddings with LLM word embeddings by selecting descriptive words as prototypes and using contrastive learning to increase similarity with time series patterns. Similarly, S2IP-LLM~\cite{pan2024s2ip} improves the LLM comprehension of time series data through semantic anchor prompts, selecting anchors based on the similarity between word embeddings and time series embeddings. In addition, \cite{liu2024calf} combines reduced-dimensionality vocabulary embeddings with cross-attention for better alignment, alongside feature regularization loss. To investigate LLMs' capability in understanding various types of temporal sequences, \cite{tang2024time} conducted experiments which indicate that LLMs perform better at predicting sequences exhibiting clear trends and seasonal characteristics.

iii) \emph{LLMs as predictor}. These works delves into the potential of directly employing LLMs for time series/energy forecasting. By meticulously crafting numerical data tokenization encoding methods that are compatible with various LLMs, this work enables the models to accurately understand energy data, thus facilitating the forecasting of future distributions under zero-shot conditions~\cite{gruver2023large}. In addition, \cite{xue2023utilizing} employs  prompt-based techniques to incorporate energy data into LLMs, fine-tuning them for autoregressive predictions.

\noindent \textbf{The future of energy management.} LLMs can enhance energy management by improving prediction interpretability and addressing complex energy allocation with agent frameworks. With advanced NLP capabilities, they can convert forecasts into human-readable explanations, helping stakeholders understand prediction rationales. This includes using techniques like prompt engineering and few-shot learning for interpretable energy usage forecasts. Additionally, LLMs can serve as intelligent agents in optimizing energy distribution, addressing demand fluctuations and renewable generation variability. Using API tools for data collection and COT analysis, they devise strategies for control and allocation, improving grid efficiency and resilience. 
Their reasoning and adaptability make them essential for managing modern power systems.

\begin{table*}[t]
\renewcommand\arraystretch{1.2}
    \centering
    \small
    \caption{A summary of LLM applications across various urban scenarios, categorized by their specific roles and task types.}
    \vspace{-0.05in}
    \label{tab:categories}
    \scalebox{0.52}{
    \begin{tabular}{c c l}
        \hline
        \multirow{1}*{\textbf{Domain}} & \textbf{Roles of LLMs} & \multicolumn{1}{c}{\textbf{Application: Work (Source) References}}\\
        \hline
        \multirow{9}*{Transportation}  
            & \multirow{2}*{Encoder} 
                  & \textbf{Traffic prediction:} ST-LLM (MDM 2024)~\cite{liu2024spatial_STLLM};  TPLLM (arXiv 2024)~\cite{ren2024TPLLM}; STD-PLM (arXiv 2024)~\cite{huang2024STDPLM}; STTLM (Sensors 2024)~\cite{ma2024spatial_STTLM}\\
                  & & \textbf{Traffic imputation:} GATGPT (arXiv 2023)~\cite{chen2023GATGPT}; STLLM-DF (arXiv 2024)~\cite{shao2024STLLMDF}\\
                  \cline{3-3}
            & \multirow{1}*{Enhancer} 
                  & \textbf{Traffic prediction:} STG-LLM (arXiv 2024)~\cite{liu2024how_STGLLM}; STGCN-L (arXiv 2024)~\cite{li2024spatio_STGCNL}.
                  \textbf{Delivery Demand prediction:} IMPEL (arXiv 2024)~\cite{nie2024joint_IMPEL}\\
            & \multirow{1}*{\&} 
                  & \textbf{Traffic signal control:} PromptGAT (AAAI 2024)~\cite{da2024prompt_PromptGAT}; iLLM-TSC (arXiv 2024)~\cite{pang2024illmtsc}; LLMlight (arXiv 2024)~\cite{lai2024llmlight}\\
            & \multirow{1}*{Encoder} 
                  & \textbf{Diverse Traffic management tasks:} TransGPT (arXiv 2024)~\cite{wang2024transgpt}. \textbf{Traffic imputation:} GT-TDI (Inf Fusion 2024)~\cite{zhang2024semantic_GT-TDI}\\
                  \cline{3-3}
            & \multirow{1}*{Predictor} 
                  & \textbf{Traffic prediction:} UrbanGPT (KDD 2024)~\cite{li2024URBANGPT};  xTP-LLM (arXiv 2024)~\cite{guo2024towards_XTPLLM}\\
                  \cline{3-3}
            & \multirow{2}*{Agent} 
                  & \textbf{Diverse Traffic management tasks:} OpenTI (arXiv 2023)~\cite{da2023openti};  TrafficGPT (Transport Policy 2024)~\cite{zhang2024TrafficGPT}; TP-GPT (arXiv 2024)~\cite{wang2024traffic_TPGPT}\\
                  & & \textbf{Traffic signal control:} LA-Light (arXiv 2024)~\cite{wang2024llm_LAlight}\\
                  \cline{3-3}
            & \multirow{1}*{Assistant} 
                  & \textbf{Traffic signal control:} Villarreal \etal (ITSC 2023)~\cite{villarreal2023can};  Tang \etal (ANZCC 2024)~\cite{tang2024large}; Dai \etal (RFID 2024)~\cite{dai2024large}; Masri \etal (arXiv 2024)~\cite{masri2024leveraging}\\
                  
        \cline{1-3}
        \multirow{6}*{Public safety}  
            & \multirow{1}*{Encoder} 
                  & \textbf{Traffic accident classification:} Grigorev \etal (arXiv 2024)~\cite{grigorev2024enhancing}\\
                  \cline{3-3}
            & \multirow{1}*{Predictor} 
                  & \textbf{Crime prediction:} UrbanGPT (KDD 2024)~\cite{li2024URBANGPT}; Sarzaeim \etal (CANAI 2024)~\cite{sarzaeim2024experimental}.\\
                  \cline{3-3}            
            & \multirow{4}*{Assistant} 
                  & \textbf{Traffic safety:} Zheng \etal (arXiv 2023)~\cite{zheng2023chatgpt};  Trafficsafetygpt (arXiv 2023)~\cite{zheng2023trafficsafetygpt}; Zarzà \etal (Sensors 2023)~\cite{Zarzà2023llm}\\
                  & & \textbf{Crime monitoring:} Watchovergpt (COMPSAC 2024)~\cite{shahid2024WatchOverGPT}.
                  \textbf{Fire engineering:} Hostetter \etal (Nat Hazards 2024)~\cite{hostetter2024the}\\
                  & & \textbf{Traffic accident information extraction:} Mumtarin \etal (arXiv 2023)~\cite{mumtarin2023large}; Zhen \etal arXiv 2024)~\cite{zhen2024leveraging}; Zhou \etal arXiv 2024)~\cite{zhou2024gpt4v}\\
                  & & \textbf{Emergency management:} Chen \etal (Int J Disast Risk Re 2024)~\cite{chen2024enhancing}; Otal \etal (CAI 2024)~\cite{otal2024llmassisted}\\
                  
        \cline{1-3}
        \multirow{7}*{Urban Mobility}  
            & \multirow{1}*{Encoder} 
                  & \textbf{Trajectory modeling:} PLMTraj (arXiv 2024)~\cite{zhou2024plm4traj}; Liu \etal (arXiv 2024)~\cite{liu2024semantic}\\
            & \multirow{1}*{\&} 
                  & \textbf{POIs information encoding:} POIGPT (ISPRS Archives 2024)~\cite{kim2024poi}; M3PT (KDD 2023)~\cite{yang2023m3pt}; LARR (Recsys 2024)~\cite{wan2024larr}\\
            & \multirow{1}*{Enhancer} 
                  & \textbf{Multimodal geographic information understanding:} ReFound (KDD 2024)~\cite{xiao2024refound}; CityGPT (arXiv 2024)~\cite{feng2024citygpt}\\
                  \cline{3-3}
            & \multirow{2}*{Predictor}
                  & \textbf{POIs recommendation:} LLM-Mob (arXiv 2023)~\cite{wang2023would}; LLMMove (CAI 2024)~\cite{feng2024move}; Beneduce \etal (arXiv 2024)~\cite{beneduce2024large}; LAMP (arXiv 2024)~\cite{balsebre2024lamp}; \\
                  & & LLM-MPE (Comput Environ Urban Syst 2023)~\cite{liang2024exploring}; LLM4POI (SIGIR 2024)~\cite{li2024large}. \textbf{OD flow prediction:} LLM-COD (arXiv 2023)~\cite{yu2024harnessing}\\
                  \cline{3-3}
            & \multirow{2}*{Agent}
                  & \textbf{Mobility generation:} AgentMove (arXiv 2024)~\cite{feng2024agentmove}; CoPB (arXiv 2024)~\cite{shao2024chain};\\
                  & & MobAgent (arXiv 2024)~\cite{li2024more}; LLMob (arXiv 2024)~\cite{wang2024large}; MobilityGPT (arXiv 2024)~\cite{haydari2024mobilitygpt}\\

        \cline{1-3}
        \multirow{7}*{Environment}
            & \multirow{1}*{Predictor} 
                  & \textbf{Weather forecasting:} PromptCast: (TKDE 2023)~\cite{Xue2024PromptCast}. \textbf{Wind speed forecasting:} STELLM (Applied Energy 2024)~\cite{wu2024STELLM}\\
                  \cline{3-3}
            & \multirow{3}*{Assistant} 
                  & \textbf{Typhoon disaster QA:} Xia \etal: (ISPRS Int. J. Geoinf 2024)~\cite{xia2024question}. \textbf{Flood disaster QA:} Sun \etal: (ICAICE 2023)~\cite{sun2023unleashing}\\
                  & & \textbf{Climate related tasks:} Arabic Mini-ClimateGPT: (EMNLP 2023)~\cite{mullappilly2023arabic}; ClimateBert (arXiv 2022)~\cite{leippold2022climatebert}; Vaghefi \etal (NeurIPS workshop 2022)~\cite{vaghefi2022deep}\\
                  & & \textbf{Earthquake impact analysis:} QuakeBERT (IJDRR 2024)~\cite{han2024enhanced}\\
                  \cline{3-3}
            & \multirow{2}*{Enhancer} 
                  & \textbf{Flood prediction:} Zhu \etal (IJGIS 2024)~\cite{zhu2024flood}. \textbf{Climate related tasks:} ChatClimate (Commun. Earth Environ. 2023)~\cite{vaghefi2023chatclimate};\\
                  & & \textbf{Fire safety planning:} Durmus \etal (ISARC 2024)~\cite{durmus2024role}. \textbf{Earthquake damage estimation:} Wang \etal (IJDRR 2024)~\cite{wang2024Near}\\
                  \cline{3-3}
            & \multirow{1}*{Agent} 
                  & \textbf{Climate related tasks:} Kraus \etal (arXiv 2023)~\cite{kraus2023enhancing}\\

        \cline{1-3}
        \multirow{4}*{Travel planning}
            & \multirow{1}*{Encoder} 
                  & \textbf{Travel planning:} Meyer \etal (arXiv 2024)~\cite{meyer2024acomparison}; \textbf{Tourist attraction description:} TourLLM (arXiv 2024)~\cite{wei2024tourllm}\\
                  \cline{3-3}
            & \multirow{1}*{Predictor} 
                  & \textbf{Travel needs extraction:} Barandon \etal (arXiv 2024)~\cite{barandoni2024automating}; 
                  \textbf{Transportation choice prediction:} Zhai \etal (arXiv 2024)~\cite{zhai2024enhancing}; Mo \etal (arXiv 2023)~\cite{mo2023large};\\
                  \cline{3-3}
            & \multirow{2}*{Agent}
                  & \textbf{Travel planning benchmark:} TravelPlanner (ICML 2024)~\cite{xie2024TravelPlanner}; \textbf{Travel planning:} TravelAgent (arXiv 2024)~\cite{chen2024travelagent};\\
                  & & Hao \etal (arXiv 2024)~\cite{hao2024large}; Li \etal (DOCS 2024)~\cite{li2024research}; ITINEAR (EMNLP 2024)~\cite{tang2024itinera}; TRIP-PAL (arXiv 2024)~\cite{rosa2024trippal}\\

        \cline{1-3}
        \multirow{2}*{Urban planning}
            & \multirow{1}*{Assistant} 
                  & \textbf{Urban planning:} HSC-GPT \etal (arXiv 2024)~\cite{ran2023hsc}; UrbanLLM (EMNLP 2024)~\cite{jiang2024urbanllm}\\
                  \cline{3-3}
            & \multirow{1}*{Agent} 
                  & \textbf{Urban planning:} Zhou \etal (arXiv 2024)~\cite{zhou2024large}; PlanGPT (arXiv 2024)~\cite{zhu2024plangpt}\\
        
        \cline{1-3}
        \multirow{4}*{Energy}
            & \multirow{1}*{Encoder} 
                  & \textbf{Energy forecasting:} Zhou \etal (NeurIPS 2023)~\cite{zhou2023ofa}; LLM4TS (arXiv 2024)~\cite{chang2024llm4ts}. \textbf{Evaluation:} Tan \etal (arXiv 2024)~\cite{tan2024are}\\
                  \cline{3-3}
            & \multirow{1}*{Encoder \&} 
                  & \textbf{Energy forecasting:} Autotimes (arXiv 2024)~\cite{liu2024autotimes}; TEMPO (ICLR 2024)~\cite{cao2024tempo}; UniTime (WWW 2024)~\cite{liu2024unitime}; TIME-FFM (NeurIPS 2024)~\cite{liu2024timeffm}\\
            & \multirow{1}*{Enhancer} 
                  & TIME-LLM (ICLR 2024)~\cite{jin2024timellm}; TEST (ICLR 2024)~\cite{sun2024test}; S2IP-LLM (arXiv 2024)~\cite{pan2024s2ip}; CALF (arXiv 2024)~\cite{liu2024calf}; \textbf{Evaluation:} Tang \etal (arXiv 2024)~\cite{tang2024time}\\     
                  \cline{3-3}
            & \multirow{1}*{Predictor} 
                  & \textbf{Energy forecasting:} Gruver \etal (NeurIPS 2023)~\cite{gruver2023large}; Xue \etal (BuildSys 2023)~\cite{xue2023utilizing}\\    
        \cline{1-3}
        \multirow{10}*{Geoscience}
            & \multirow{5}*{Precidtor} 
                  & \textbf{Geographic knowledge understanding:} GPT4Geo (arXiv 2023)~\cite{roberts2023gpt4geo}; Bhandari \etal (SIGSPATIAL 2023)~\cite{bhandari2023large}; K2 (WSDM 2024)~\cite{deng2024k2};\\
                  & & Liu \etal (AGILE: GISS 2024)~\cite{liu2024measuring}; Kopanov \etal (AICCONF 2024)~\cite{kopanov2024comparative}; GeoLLM (ICLR 2024)~\cite{manvi2024geollm}; WorldBench (ACM FAccT 2024)~\cite{moayeri2024worldbench};\\
                  & & Fernandez \etal (arXiv 2023)~\cite{fernandez2023core}; Mooney \etal (SIGSPATIAL workshop 2023)~\cite{mooney2023towards}; BB-GeoGPT \etal (Inf Process Manag 2023)~\cite{zhang2024bb}\\
                  & & \textbf{Geolocation:} LLMGEO (arXiv 2024)~\cite{wang2024llmgeo}; GeoReasoner (ICML 2024)~\cite{li2024georeasoner}; GeoLocator (Applied Sciences 2024)~\cite{yang2024geolocator}\\
                  & & \textbf{Human mobility prediction:} GEOFormer (HuMob-Challenge workshop 2023)~\cite{solatorio2023geoformer}\\
                  \cline{3-3}
            & \multirow{1}*{Enhancer} 
                  & \textbf{Geographic knowledge understanding:} GeoLM (arXiv 2023)~\cite{li2023geolm}; UrbanClip (WWW 2024)~\cite{yan2024urbanclip}.
                  \textbf{Travel query understanding:} QUERT (KDD 2023)~\cite{xie2023quert}\\
                  \cline{3-3}    
            & \multirow{4}*{Agent}
                  & \textbf{Geo benchmark:} GeoLLM-Engine (CVPR 2024)~\cite{singh2024geollm}.
                  \textbf{Geo infomation QA:} GeoQAMap (GIScience 2023)~\cite{feng2023geoqamap}; MapGPT (SIGSPATIAL 2023)~\cite{zhang2023map}.\\
                  
                  & & \textbf{Address standardization:} GeoAgent (ACL 2024)~\cite{huang2024geoagent}.
                  \textbf{socioeconomic indicator estimates:} GeoSEE (arXiv 2024)~\cite{han2024geosee}\\
                  
                  & & \textbf{Multiple geospatial tasks:} GeoGPT (Int J Appl Earth Obs Geoinf 2024)~\cite{zhang2024geogpt}. \textbf{Active geo-localization:} GOMAA-Geo (arXiv 2024)~\cite{sarkar2024gomaa}\\
                  
                  & & \textbf{Urban knowledge graph construction:} UrbanKGent (arXiv 2024)~\cite{ning2024urbankgent}\\
                  \cline{3-3} 

        \cline{1-3}
        \multirow{6}*{Autonomous Driving}
            & \multirow{1}*{Encoder} 
                  & \textbf{Autonomous driving:} RAG-Driver (arXiv 2024)~\cite{yuan2024rag}; Talk2BEV (ICRA 2024)~\cite{choudhary2023talk2bev}; Gopalkrishnan \etal (CVPR workshop 2024)~\cite{gopalkrishnan2024multi}; \\
                  \cline{3-3} 
            & \multirow{2}*{Predictor} 
                  & \textbf{Autonomous driving:} DriveVLM (arXiv 2024)~\cite{tian2024drivevlm}; DriveGPT4 (IEEE Robot. Autom. Lett 2024)~\cite{xu2024drivegpt4};\\ 
                  & & LanguageMPC (arXiv 2023)~\cite{sha2023languagempc}; AsyncDriver (ECCV 2024)~\cite{chen2025asynchronous}; LMDrive (CVPR 2024)~\cite{shao2024lmdrive};\\
                  \cline{3-3} 
            & \multirow{2}*{Agent} 
                  & \textbf{Autonomous driving:} Agent-Driver (COLM 2024)~\cite{mao2023language}; DiLu (ICLR 2024)~\cite{wen2024dilu}; DriveLikeAHuman (WACV 2024)~\cite{fu2024drive}; \\
                  & & SurrealDriver (arXiv 2024)~\cite{yang2024driving}; DriveLLM (TIV 2023)~\cite{cui2023drivellm}; Agentscodriver (arXiv 2024)~\cite{hu2024agentscodriver}\\
                  \cline{3-3} 
            & \multirow{1}*{Enhancer} 
                  & \textbf{Autonomous driving:} TARGET (arXiv 2023)~\cite{deng2023target}; ChatSim (CVPR 2024)~\cite{wei2024editable}; REvolve (arXiv 2024)~\cite{hazra2024revolve}\\
        
        \hline
    \end{tabular}
    }
    \vspace{-0.15in}
\end{table*}

\subsection{Geoscience} 
The geospatial data supports advancements in earth science, encompassing remote sensing images, urban road networks, and geographic textual materials. Utilizing this data, researchers can develop various applications such as geo-localization, geo-knowledge reasoning, and object detection.

\noindent \textbf{Core tasks \& previous works.} Geoscience tasks are diverse and can be formalized as mappings between multimodal inputs and outputs. Input modalities often encompass textual geographical context, visual information, and task-specific data. Outputs, which vary depending on the task, may consist of language responses for open-ended QA scenarios, executable code for API calls, or labels for geo-related classification tasks.

Earlier studies in Geoscience have tailored workflows or neural modules for various geospatial data modalities. For language inputs, traditional applications~\cite{liu2022geoparsing, hu2018eupeg, wang2019enhancing} employ named entity recognition tools and GIS techniques to link toponyms in text to geographic databases, using heuristic rules and deep learning. For image inputs, they often use specialized loss functions and models, as seen in geographic image location applications~\cite{vivanco2024geoclip, hays2008im2gps, wu2022im2city}.

\noindent\textbf{GeoScience Powered by LLMs.} Integrating LLMs has enabled researchers to enhance geospatial applications by transforming original data into LLM-readable content for improved reasoning and empowering LLMs with access to geo-related tools. We summarize these developments as follows:

i) \emph{LLMs as predictor}. These efforts seek to enable LLMs to reason from geospatial text inputs to address user queries, with early research exploring their geographical knowledge depth. GPT4GEO~\cite{roberts2023gpt4geo} assessed GPT-4's factual geographical knowledge and interpretative reasoning capabilities. In addition, \cite{mooney2023towards} evaluated ChatGPT's performance in GIS contexts and its comprehension of spatial concepts, while \cite{bhandari2023large} confirmed LLMs' effectiveness in handling geospatial questions, focusing on aspects such as geospatial knowledge, awareness, and reasoning skills. 
Besides, related studies have explored various tasks such as image geolocation~\cite{wang2024llmgeo,li2024georeasoner}, natural language-based geo-guessing~\cite{liu2024measuring}, and assessing GPT-4's capacity to derive geospatial data from images or social media~\cite{yang2024geolocator}. Other research focused on evaluating LLMs' performance in multilingual geo-entity detection~\cite{kopanov2024comparative} and investigating geographical biases in these models~\cite{manvi2024geollm, moayeri2024worldbench}.

Further efforts have enhanced LLM performance on geo-related tasks through strategies like fine-tuning and prompt engineering. For instance, MapGPT~\cite{fernandez2023core} employs retrieval-augmented generation (RAG) techniques to improve LLMs' understanding and responses to location-based queries. GeoFormer~\cite{solatorio2023geoformer} involves training a pre-trained language model to develop an effective model capable of predicting human mobility.
Additionally, Both K2~\cite{deng2024k2} and BB-GeoGPT~\cite{zhang2024bb} developed instruction fine-tuning databases enriched with geographical knowledge for LLMs, enhancing their ability to accurately answer geo-related queries.

ii) \emph{LLMs as enhancer}. When utilizing LLMs as enhancers, geographic data is preprocessed or pretrained to improve downstream tasks. GeoLM~\cite{li2023geolm} employs contrastive learning and masked language modeling to train on geospatial data, enabling applications in toponym recognition, geographic entity classification, and relationship extraction. QUERT~\cite{xie2023quert}, pretrained for travel query understanding, improves travel search performance through four tailored tasks. In addition, UrbanCLIP~\cite{yan2024urbanclip} integrates Image-to-Text LLMs with satellite images to create image-text pairs, enhancing urban analysis tasks such as population and GDP indicator predictions.

iii) \emph{LLMs as agent}. Agent-based methods enable LLMs to access external tools for geo-related tasks. Singh et al. introduced the GEOLLM-Engine Benchmark~\cite{singh2024geollm} to evaluate geo-copilots on complex real-world tasks. Other works focus on specific geo-tasks: GeoQAMap~\cite{feng2023geoqamap} converts GeoQA queries into SPARQL to retrieve data and generate interactive maps, while Map GPT Playground~\cite{zhang2023map} enhances geography-related queries via mapping services. GeoAgent~\cite{huang2024geoagent} employs geospatial tools for address standardization, and GeoSEE~\cite{han2024geosee} estimates socioeconomic indicators by selecting relevant modules. In addition, GeoGPT~\cite{zhang2024geogpt} integrates LLMs with GIS tools to automate geospatial tasks and improve workflows. For active geo-localization, GOMAA-GeoGOal~\cite{sarkar2024gomaa} achieves zero-shot generalization across target modalities for efficient navigation, and the UrbanKGent~\cite{ning2024urbankgent} framework constructs a rich urban knowledge graph from minimal data.

\noindent \textbf{Towards Autonomous Geo-Intelligent Systems.} The future of Geoscience lies in developing autonomous geo-intelligent systems that integrate LLMs with real-time geospatial data. This involves LLMs autonomously interacting with geospatial databases, satellite imagery, and sensory data to make informed decisions in dynamic settings. Incorporating reinforcement learning will enable the system to adapt its responses based on real-world outcomes. A feedback loop will allow continuous refinement of models, improving tasks like environmental monitoring and urban planning. Additionally, embedding ethical frameworks will ensure responsible operation of these systems, considering societal impacts and promoting equitable access to geospatial insights.

\subsection{Autonomous Driving}
Autonomous driving utilizes sensors such as cameras, radar, and LiDAR, alongside vehicle-mounted devices, to process real-time data from the vehicle's surroundings. This information enables the vehicle to detect and navigate around obstacles, recognize traffic signs, and adjust to changing road conditions, thereby enhancing driving safety and efficiency.

\noindent\textbf{Core tasks \& previous works}.
The core tasks of autonomous driving with LLMs can be formalized as a multi-modal mapping function $f: \mathcal{V} \times \mathcal{T} \rightarrow \mathcal{Y}$, where $\mathcal{V}$ represents visual inputs (images, videos, bird’s-eye view (BEV) maps), $\mathcal{T}$ represents language inputs (queries, instructions, rules), and $\mathcal{Y}$ represents the output space (scene descriptions, control signals, semantic maps). The visual input space can be further decomposed as $\mathcal{V} = {\mathcal{I}^{M \times H \times W \times 3}, \mathcal{P}^{N \times 4}, \mathcal{B}^{H' \times W' \times C}}$, representing multi-view images, point clouds, and BEV features respectively. The task aims to use LLMs to bridge the semantic gap between visual perception and high-level understanding, while maintaining computational efficiency and real-time performance.

Prior to the adoption of LLMs, autonomous driving primarily relied on three categories of approaches: (1) Rule-based systems~\cite{treiber2000congested, sun2022lawbreaker} that encode domain knowledge and traffic regulations explicitly but suffer from poor generalization to unseen scenarios; (2) Traditional computer vision pipelines~\cite{li2022bevformer, wang2020v2vnet} that decompose the task into sequential modules (detection, segmentation, tracking), often using CNNs or transformers as backbone networks; (3)  End-to-end learning approaches~\cite{hu2023planning, shao2023safety} that directly map visual inputs to desired outputs through deep neural networks, though often lacking interpretability. These methods typically optimize objectives like $\min_\theta \mathcal{L}(\hat{y}, y)$ where $\hat{y} = f_\theta(\mathcal{V})$, focusing solely on the visual domain without incorporating language understanding. They face challenges in handling complex scenarios, long-tail cases, and providing natural language explanations - limitations that motivated the integration of LLMs for enhanced reasoning and interpretation capabilities.

\noindent\textbf{Leveraging LLMs for Autonomous Driving.}
The integration of large language models into autonomous driving marks a significant advancement by bridging visual perception with semantic understanding and enabling better handling of complex scenarios through common sense reasoning. LLMs can serve multiple roles in this domain:

i) \emph{LLMs as encoder}.
Recent works have explored using LLMs as encoders to transform complex urban scene information into meaningful representations. RAG-Driver~\cite{yuan2024rag} introduces a novel retrieval-augmented approach, where LLMs not only encode current scenes but also facilitate experience retrieval for better generalization, while Talk2BEV~\cite{choudhary2023talk2bev} demonstrates LLMs' capability in spatial reasoning through BEV representation encoding. In addition, a method for autonomous driving visual question answering is proposed, leveraging a visual encoder and mapping layer to align image-text representations and fine-tuning the T5 model with instruction tuning for task completion~\cite{gopalkrishnan2024multi}. These works share a common emphasis on efficient feature fusion across multiple modalities (visual, spatial, and temporal) while addressing the challenges of real-time processing, showing promise in bridging the gap between raw sensory inputs and high-level semantic understanding required for autonomous driving systems.

ii) \emph{LLMs as predictor}.
LLMs have emerged as powerful predictors in autonomous driving systems, focusing on translating high-level understanding into concrete driving decisions and control signals. 
DriveVLM~\cite{tian2024drivevlm} expands this by incorporating BEV for more comprehensive scene understanding. In addition,
DriveGPT4~\cite{xu2024drivegpt4} and LMDrive~\cite{shao2024lmdrive} demonstrate end-to-end approaches where LLMs directly predict control signals from video and multi-modal sensor inputs respectively, while incorporating natural language explanations for interpretability. LanguageMPC~\cite{sha2023languagempc} takes a hybrid approach by using LLMs to make high-level decisions that are then translated into Model Predictive Control parameters, bridging the gap between linguistic reasoning and traditional control methods. AsyncDriver~\cite{chen2025asynchronous} focuses on efficient planning decisions, introducing asynchronous processing to balance computational costs with real-time requirements. Talk2BEV~\cite{choudhary2023talk2bev} extends these capabilities to BEV representations, generating decisions based on a more comprehensive spatial understanding. These methods leverage LLMs' reasoning capabilities to address challenges in real-time processing, safety constraints, and translating linguistic decisions to control signals.

iii) \emph{LLMs as agent}.
Several works have explored using LLMs as autonomous agents in driving scenarios, focusing on human-like decision-making. Agent-Driver~\cite{mao2023language} and DiLu~\cite{wen2024dilu} pioneered this approach by implementing LLMs as cognitive agents with tool libraries and memory mechanisms, while DriveLikeAHuman~\cite{fu2024drive} and SurrealDriver~\cite{yang2024driving} enhanced this concept by incorporating human driving patterns and thought processes. DriveLLM~\cite{cui2023drivellm} introduced cyber-physical feedback to enhance safety, and Agentscodriver~\cite{hu2024agentscodriver} extended the paradigm to multi-vehicle collaboration. RAG-Driver~\cite{yuan2024rag} further advanced the field by introducing RAG for better experiential learning. These approaches share common themes of combining LLMs' reasoning capabilities with domain-specific knowledge and continuous learning mechanisms, addressing challenges in real-time decision-making, safety constraints, and human-like behavior generation.

iv) \emph{LLMs as enhancer}.
Research has demonstrated LLMs' capability to enhance various components of autonomous driving systems beyond direct control or decision-making. TARGET~\cite{deng2023target} leverages LLMs to automatically translate natural language traffic rules into executable test scenarios, reducing manual effort and ensuring comprehensive testing coverage. ChatSim~\cite{wei2024editable} enhances simulation by enabling natural language-based scene editing and asset integration, making simulation environments more flexible and user-friendly. REvolve~\cite{hazra2024revolve} takes a unique approach by using LLMs to evolve reward functions based on human feedback, bridging the gap between human knowledge and numerical optimization objectives. These approaches demonstrate LLMs' potential in enhancing various aspects of autonomous driving development, from testing and simulation to reward design, while maintaining interpretability and user accessibility.

\noindent\textbf{Towards Collaborative LLM Agents for Autonomous Driving}.
A promising direction in autonomous driving involves developing collaborative multi-agent frameworks that utilize LLMs in various interconnected roles within a unified system. This approach would allow LLMs to function not just as isolated components but as a collective network that communicates and reasons together, enhancing robustness and adaptability in urban analysis systems. 
Future research should focus on developing efficient multi-agent coordination algorithms, scalable architectures for real-time operations, and safety, interpretability, and compliance mechanisms. Additionally, integrating domain-specific knowledge and ethical considerations into these LLM-driven systems could improve their effectiveness and alignment with human values, advancing intelligent transportation solutions.

%% file: UCLLM_data.tex
\section{Evaluation and Data Resources}
\label{sec:resources}

\begin{table*}[t]
\renewcommand\arraystretch{1.2}
    \centering
    \small
    \caption{Summary of commonly used datasets across various urban domains: data types, dataset names, links, sources, and coverage.}
    \vspace{-0.05in}
    \label{tab:data}
    \scalebox{0.51}{
    \begin{tabular}{|c|c|c|c|c|}
        \hline
        \multirow{1}*{\textbf{Domain}} & \textbf{Data types} & \textbf{Dataset (link) \& References} & \textbf{Source (link)} & \textbf{Region}\\
        \hline
        \multirow{10}*{Transportation}  
            & \multirow{1}*{Traffic flow} 
                  & \href{https://github.com/Davidham3/STSGCN}{PEMS0X}~\cite{song2020STSGCN}; \href{https://github.com/liuxu77/LargeST}{LargeST}~\cite{liu2023largest} & \href{https://pems.dot.ca.gov/}{Caltrans PEMS} & California (CA)\\
                  \cline{2-5}
            & \multirow{3}*{Traffic speed}
                  & \href{https://github.com/VeritasYin/STGCN_IJCAI-18}{PEMS07(M/L)}~\cite{yu2018spatio_STGCN}; \href{https://github.com/liyaguang/DCRNN}{PEMS-BAY, METR-LA}~\cite{li2018diffusion_DCRNN}  & \href{https://pems.dot.ca.gov/}{Caltrans PEMS}; \href{https://www.metro.net/}{Metro} & CA; SF Bay Area; LA County \\
                  \cline{3-5}
                & & \href{https://github.com/JingqingZ/BaiduTraffic}{Q-Traffic}~\cite{liao2018qtraffic}; \href{https://github.com/tsinghua-fib-lab/UniST}{TrafficXX}~\cite{yuan2024unist}; \href{https://github.com/lehaifeng/T-GCN/tree/master/data}{SZ-TAXI}~\cite{li2020tgcn} & N.A. & China (\eg~Beijing, Shenzhen)\\
                \cline{3-5}
                & & \href{https://github.com/zhiyongc/Seattle-Loop-Data}{LOOP-SEATTLE}~\cite{cui2020loopseattle}; \href{https://github.com/RomainLITUD/Multistep-Traffic-Forecasting-by-Dynamic-Graph-Convolution}{ROTTERDAM}~\cite{li2021rotterdam} & N.A. & Greater Seattle Area; Rotterdam\\
                \cline{2-5}
            & \multirow{1}*{Traffic index}
                  & \href{https://github.com/RobinLu1209/ST-GFSL}{Chengdu-didi, Shenzhen-didi}~\cite{lu2022didi}  & N.A & Chengdu; Shenzhen\\
                  \cline{2-5}
            & \multirow{1}*{People flow}
                  & \href{https://github.com/TolicWang/DeepST/tree/master/data/TaxiBJ}{TaxiBJ}~\cite{zhang2017taxibj}; \href{https://github.com/JinleiZhangBJTU/ResNet-LSTM-GCN}{BJ-SW}~\cite{zhang2021beijingsubway}; \href{https://github.com/ivechan/PVCGN}{HZ,SHMETRO}~\cite{liu2022metro}  & N.A & Beijing; Hangzhou; Shanghai\\
                  \cline{2-5}
            & \multirow{1}*{Taxi}
                  & \href{https://www.microsoft.com/en-us/research/publication/t-drive-trajectory-data-sample/}{T-Drive}~\cite{yuan2010tdrive}; \href{https://huggingface.co/datasets/bjdwh/UrbanGPT_ori_stdata/tree/main}{NYC-TAXI; CHI-TAXI}~\cite{li2024URBANGPT}  & \href{https://opendata.cityofnewyork.us/}{NYC OpenData};\href{https://data.cityofchicago.org/}{CHI Data Portal}  & Beijing; New York; Chicago \\
                  \cline{2-3} \cline{5-5}
            & \multirow{1}*{Bike}
                  & \href{https://github.com/KL4805/CrossTReS}{NYC-BIKE, CHI-BIKE, DC-BIKE}~\cite{jin2022bike}  & \href{https://citibikenyc.com/system-data}{citibike}; \href{https://divvybikes.com/system-data}{divvybike}; \href{https://capitalbikeshare.com/system-data}{capitalbike} & New York; Chicago; Washington DC\\
                  \cline{2-5}
            & \multirow{2}*{Traffic signal}
                  & \href{https://github.com/wingsweihua/colight}{JN-SN, HZ-SN}~\cite{wei2019colight} & N.A. & Jinan; Hangzhou\\
                  \cline{3-5}
                  & & \href{https://github.com/Chacha-Chen/MPLight}{Manhattan-SN}~\cite{chen2020toward}; \href{https://github.com/wingsweihua/colight}{NYC-SN}~\cite{wei2019colight} & \href{https://www.openstreetmap.org/}{OpenStreetMap}; NYC \href{https://www.nyc.gov/html/dot/html/infrastructure/signals.shtml}{Sig}, \href{https://www.nyc.gov/site/tlc/about/tlc-trip-record-data.page}{Trip} & Manhattan; New York\\
                  \cline{1-5}
        \multirow{2}*{Public Safety}  
            & \multirow{1}*{Crime} 
                  & \href{https://github.com/LZH-YS1998/STHSL}{NYC-Crime, CHI-Crime}~\cite{li2022spatial_STHSL}; SF-Crime~\cite{jin2022sfcrime} & \href{https://opendata.cityofnewyork.us/}{NYC OpenData}; & New York; \\
                  \cline{2-3}
            & \multirow{1}*{Traffic accident} 
                  & NYC-accident, CHI-accident~\cite{wang2021accident}; SF-accident & \href{https://data.cityofchicago.org/}{CHI Data Portal}; \href{https://datasf.org/opendata/}{DataSF} & Chicago; San Francisco\\
                  \cline{1-5}

        \multirow{6}*{Urban Mobility} & \multirow{4}*{POI check-in} & \href{https://sites.google.com/site/yangdingqi/home/foursquare-dataset}{Foursquare-NYC, Foursquare-Tokyo}~\cite{yang2014modeling} & \href{https://foursquare.com/}{Foursquare Platform} & New York; Tokyo \\
        \cline{3-5}
        & & \href{https://snap.stanford.edu/data/loc-gowalla.html}{Gowalla}~\cite{cho2011friendship} & \href{https://www.gowalla.com/}{Gowalla Platform} & California; Nevada\\
        \cline{3-5}
        & & \href{https://www.yelp.com/dataset/download}{Yelp}~\cite{wang2019kgat} & \href{https://www.yelp.com/}{Yelp Platform} & Around the world\\
        \cline{3-5}
        & & MPTD1, MPTD2~\cite{yang2023m3pt} & \href{https://www.fliggy.com/}{Fliggy Platform} & Around the world\\
        \cline{2-5}
        
        & \multirow{2}*{GPS trajectory} & \href{https://www.microsoft.com/en-us/research/publication/geolife-gps-trajectory-dataset-user-guide/}{Geolife}~\cite{zheng2010geolife} & \href{https://www.microsoft.com/en-us/research/lab/microsoft-research-asia/}{Microsoft Research Asia} & Beijing\\
        \cline{3-5}
        & & \href{https://archive.ics.uci.edu/dataset/339/taxi+service+trajectory+prediction+challenge+ecml+pkdd+2015}{Portal-taxi}~\cite{jiang2023continuous} & N.A. & Porto\\
        \hline
        
        \multirow{7}*{Environmental Monitoring}  
            & \multirow{3}*{Weather Condition} 
                  & \href{https://github.com/zhouhaoyi/Informer2020}{Weather}~\cite{zhou2021informer} & \href{https://www.ncei.noaa.gov/data/local-climatological-data/}{NCEI} & USA\\
                  \cline{3-5}
                  & & \href{https://github.com/HaoUNSW/PISA}{City Temperature (CT)}~\cite{Xue2024PromptCast} & University of Dayton & Around the world \\
                  \cline{3-5}
                  & & DSWE1, DSWE2, NREL, SDWPF~\cite{wu2024STELLM} & N.A. & USA \\
                  \cline{2-5}
            & \multirow{1}*{Disastrous Incident} 
                  & \href{https://github.com/SusuXu-s-Lab/Hierarchical-Earthquake-Casualty-Information-Retrieval}{Global Earthquake Event}~\cite{wang2024Near} &  \href{https://earthquake.usgs.gov/data/pager/}{PAGER system} & Around the world \\
                  \cline{2-5}
            & \multirow{3}*{Climate} 
                  & Clima500-Instruct~\cite{mullappilly2023arabic} & \href{https://github.com/SmartDataAnalytics/Climate-Bot}{CCMRC}, Clima-QA & Around the world \\
                  \cline{3-5}
                  & & FFD-IQA~\cite{sun2023unleashing} & \href{https://crisisnlp.qcri.org/crisismmd.html}{CrisisMMD}, \href{https://github.com/BinaLab/FloodNet-Challenge-EARTHVISION2021}{FloodNet}, \href{https://github.com/cvjena/eu-flood-dataset}{European Flood 2013 Dataset} & Around the world \\
                  \cline{3-5}
                  & & \href{https://github.com/tdiggelm/climate-fever-dataset}{Climate Fever dataset}~\cite{vaghefi2022deep} & N.A. & Around the world \\
                  \cline{1-5}
        \multirow{2}*{Travel}  
            & \multirow{1}*{QA \& Search } 
                  & \href{https://github.com/soniawmeyer/WanderChat}{WanderChat}~\cite{meyer2024acomparison}; \href{https://github.com/hsaest/QUERT}{QUERT}~\cite{xie2023quert} & \href{https://developers.reddit.com/docs/api}{Reddit}; \href{https://open.fliggy.com}{Fliggy}. & Around the world; China \\
                  \cline{2-5}
            & \multirow{1}*{Planning} 
                  & \href{https://github.com/OSU-NLP-Group/TravelPlanner}{TravelPlanner}~\cite{xie2024TravelPlanner}; \href{https://github.com/YihongT/ITINERA}{ITINERA}~\cite{tang2024itinera} & \href{https://developers.google.com/maps/documentation}{Google Maps}, \href{https://www.kaggle.com/datasets}{Kaggle}; \href{https://lbs.amap.com/}{Amap}  & Around the world; China \\
                  \cline{1-5}
        \multirow{3}*{Urban Planning}  
            & \multirow{2}*{Urban Planning QA} 
                  & \href{https://anonymous.4open.science/r/UrbanLLM-1227/README.md}{Singapore Human-Annotated Dataset}~\cite{jiang2024urbanllm} & Singapore Open Data API & Singapore \\
                  \cline{3-5}
                  & & Urban-planning-annotation~\cite{zhu2024plangpt} & N.A. & China \\
                  \cline{2-5}
            & \multirow{1}*{Participatory Urban Planning} 
                  & HLG, DHM~\cite{zhou2024large} & N.A. & Beijing, China \\
                  
                  \cline{1-5}
        \multirow{2}*{Energy}  
            & \multirow{1}*{Electricity} 
                  & \href{https://github.com/zhouhaoyi/Informer2020}{ETT-X}~\cite{zhou2021informer}; \href{https://github.com/laiguokun/multivariate-time-series-data}{Electricity}~\cite{lai2018solarenergy} & \href{https://github.com/zhouhaoyi/ETDataset}{ETT}; \href{https://archive.ics.uci.edu/dataset/321/electricityloaddiagrams20112014}{UCI dataset} & China; USA \\
                  \cline{2-5}
            & \multirow{1}*{Solar Energy} 
                  & \href{https://github.com/laiguokun/multivariate-time-series-data}{Solar-Energy}~\cite{lai2018solarenergy} & \href{https://www.nrel.gov/grid/solar-power-data.html}{Nrel} & USA \\
                  \cline{2-3}
                  \cline{2-5} 
        \hline

        \multirow{4}*{GeoScience}
            & \multirow{1}*{Real-world Tasks}
                & GeoLLM-Engine Benchmark~\cite{singh2024geollm} & N.A. & N.A.\\
                \cline{2-5}
            & \multirow{1}*{Geolocalization}
                & \href{https://github.com/yeyimilk/LLMGeo}{LLMGeo-Benchmark}~\cite{wang2024llmgeo} & Google Street View & Around the world\\
                \cline{2-5}
            & \multirow{2}*{Documents \& Instruction \& QA}
                & \href{https://github.com/davendw49/k2}{GeoSignal/Bench}~\cite{deng2024k2} & Geoscience literatures, Wikipedia and NPEE & N.A.\\
                \cline{3-5}
                & & \href{https://github.com/AGI-GIS/BB-GeoGPT}{BB-GeoSFT/GeoPT}~\cite{zhang2024bb} & GIS-related literature and Wikipedia & N.A. \\
        \hline
        \multirow{7}*{Autonomous Driving} & \multirow{3}*{Real-world Visual Data} & \href{https://www.nuscenes.org/}{nuScenes}~\cite{caesar2020nuscenes}, \href{https://www.nuscenes.org/nuplan}{nuPlan}~\cite{caesar2021nuplan} & \href{https://www.nuscenes.org/}{nuScenes} & N.A.\\
        \cline{3-5}
        & & \href{https://waymo.com/open/download}{WOMD}~\cite{Ettinger_2021_ICCV} & \href{https://waymo.com/open/}{Waymo} & N.A.\\
        \cline{3-5}
        & & \href{https://uwaterloo.ca/watonobus/downloads}{WATonoBus}~\cite{bhatt2024watonobus} & \href{https://uwaterloo.ca/watonobus/}{WATonoBus Project} & University of Waterloo campus\\
        \cline{2-5}
        & \multirow{2}*{Simulation Data} & \href{https://github.com/carla-simulator/carla}{CARLA}~\cite{dosovitskiy2017carla}, \href{https://github.com/eleurent/highway-env}{HighwayEnv}~\cite{highway-env}, \href{https://github.com/liuyuqi123/ComplexUrbanScenarios}{IdSim}~\cite{liu2021reinforcement} & N.A. & N.A.\\
        \cline{3-5}
        & & \href{https://github.com/microsoft/airsim/releases}{AirSim}~\cite{shah2018airsim} & Microsoft & N.A. \\
        \cline{2-5}

        & \multirow{2}*{Instruction \& QA} & \href{https://llmbev.github.io/talk2bev/}{Talk2BEV-Bench}~\cite{choudhary2023talk2bev}, \href{https://github.com/JinkyuKimUCB/explainable-deep-driving}{BDD-X}~\cite{kim2018textual} & N.A. & N.A.\\
        \cline{3-5}
        & & \href{http://macchina-ai.eu/}{Talk2Car}~\cite{deruyttere2019talk2car}, \href{https://github.com/xmed-lab/NuInstruct}{NuInstruct}~\cite{ding2024holistic}, \href{https://github.com/OpenDriveLab/DriveLM}{DriveLM}~\cite{sima2023drivelm} & N.A. & N.A. \\
        \cline{3-5}
        \hline
    \end{tabular}
    }
    \vspace{-0.15in}
\end{table*}

This section compiled evaluation metrics and commonly used datasets from various domains to assist researchers in quickly understanding evaluation methods and data acquisition channels across different fields. The dataset information includes data types, data links, references, source links, and geographical coverage, as shown in Table~\ref{tab:data}.

\subsection{Transportation}
Traffic prediction involves processing a variety of data types, such as traffic flow, crowd flow, taxi trips. Data are primarily  sourced from open traffic systems (\eg~PEMS, NYC open data) and researcher-released datasets (\eg~TaxiBJ, T-Drive). Prediction accuracy is evaluated using metrics like MAE, RMSE, and MAPE~\cite{wu2019graph_GWN,bai2020adaptive_AGCRN,liu2024spatial_STLLM}.

In traffic management, the evaluation data are categorized into synthetic and real-world data. Synthetic data are created in simulation environments that replicate various traffic scenarios, including multiple intersections and vehicle flow patterns. Real-world data (\eg~Manhattan-SN, NYC-SN), on the other hand, are collected from transportation departments and map APIs, involving traffic signal configurations, city maps, and actual travel trajectories. The evaluation metrics of TSC include average traveling time, average queue length, and throughput~\cite{chen2020toward,lai2024llmlight,wang2024unitsa}.

\subsection{Public Safety}
Traffic accident prediction, severity analysis, and crime prediction are crucial for urban safety. These data often originate from open government portals, such as those in New York City and Chicago. Metrics like accuracy, precision, recall, and F1 scores are commonly used to evaluate event data~\cite{li2022spatial_STHSL,li2024URBANGPT}. LLMs are emerging as powerful tools for public safety analysis, with their effectiveness evaluated based on coverage, fluency, completeness, and accuracy. Text quality of LLM-generated content is measured using metrics such as BLEU, ROUGE and BLEURT~\cite{otal2024llmassisted,zheng2023trafficsafetygpt}.

\subsection{Urban Mobility}
Urban mobility prediction research relies on two main types of datasets: POI check-in data and GPS trajectory data. POI check-in datasets, such as Foursquare data from cities like New York and Tokyo, include user IDs, locations, timestamps, and venue categories, spanning months to years. GPS trajectory datasets, like Geolife and taxi/ride-sharing data from cities such as Beijing and Chengdu, offer granular mobility information with precise coordinates and timestamps, often containing millions of records from hundreds to thousands of users.

Evaluation metrics in this field fall into three categories: (1) Accuracy-based metrics, including Top-k Accuracy for prediction within top-k candidates~\cite{beneduce2024large}, and RMSE, MAE, and MAPE for quantifying prediction errors~\cite{liang2024exploring}; (2) Ranking-based metrics, such as Mean Reciprocal Rank (MRR)~\cite{feng2024move} and Normalized Discounted Cumulative Gain (NDCG) for ranking quality~\cite{feng2024agentmove}; (3) Distribution-based metrics, which use Jensen-Shannon divergence (JSD) to compare the similarity between predicted and actual mobility patterns in terms of step distance, and temporal intervals~\cite{li2024more}.

\subsection{Environment}
Environmental monitoring involves processing various data, including weather conditions, disasters, and climate changes, sourced from global databases and research institutions. For instance, the Weather dataset from National Centers for Environmental Information (NCEI) is widely used in forecasting tasks~\cite{zhou2021informer,zhou2022fedformer}. Metrics such as MAE, RMSE~\cite{Xue2024PromptCast,zhou2022fedformer}, accuracy, and F1 score~\cite{leippold2022climatebert} are common for prediction tasks, while ROUGE-N and Win Rate~\cite{xia2024question} are used for language tasks like summarization and information extraction.

\subsection{Travel}
The emergence of LLMs enables personalized and automated travel planning. Existing works evaluate LLM-generated plans using metrics like ROUGE, BLEU, and BERTScore~\cite{wei2024tourllm,barandoni2024automating} to compare generated content with references, or by using GPT-4 and human evaluators for scoring. In the travel planning task, TravelPlanner benchmark~\cite{xie2024TravelPlanner} introduces a dataset and three metrics: 1) Delivery Rate, measuring if a plan is produced on time; 2) Constraint Pass Rate, assessing adherence to user specifications; and 3) Final Pass Rate, evaluating whether plans fully meet user requirements. These datasets and metrics advance the field's development.

\subsection{Urban Planning and Development}
The availability of massive data has enabled the exploration of using LLMs to advance urban planning. For Urban Planning QA, ~\cite{jiang2024urbanllm} provides Singapore-specific QA pairs, while~\cite{zhu2024plangpt} offers data support in China. In participatory urban planning, datasets like HLG and DHM~\cite{zhou2024large} cover Beijing and other regions in China. These datasets provide essential resources for research and the training of LLMs in urban planning.
Evaluating LLMs in urban planning involves various metrics. For deterministic tasks like information extraction and text evaluation, accuracy and F1 score~\cite{jiang2024urbanllm,zhu2024plangpt,sideris2019using} are common, while open-ended QA tasks prioritize metrics assessing relevance, completeness, and practicality~\cite{ran2023hsc}. These metrics ensure model quality, reliability, and efficacy, helping researchers refine LLM applications in urban planning.

\subsection{Energy}
Energy management involves tasks such as electricity and energy forecasting, which are treated as time series data and analyzed using time series prediction models. Common evaluation metrics for these models include MAE and RMSE~\cite{xue2023utilizing,zhou2021informer}. Given the substantial scale variations across energy datasets, existing studies often employ normalized values to calculate these metrics. Popular datasets used in this field include ETT-X, Electricity, and Solar-Energy.

\subsection{Geoscience}
Due to the diversity of application scenarios in geo-related applications, various metrics are employed to evaluate performance across different tasks. For instance, image geolocation uses accuracy as a metric, object detection utilizes the F1 score\cite{li2024georeasoner}, question-answering tasks are assessed using the ROUGE score~\cite{singh2024geollm}, and the execution of specific tasks by agents is evaluated based on the success rate~\cite{sarkar2024gomaa}. Popular benchmarks for evaluating the performance of LLMs in this field include the GeoLLM-Engine Benchmark~\cite{singh2024geollm}, LLMGeo-Benchmark~\cite{wang2024llmgeo}, and GeoBench~\cite{deng2024k2}. Additionally, there are instruction tuning datasets such as GeoSignal~\cite{deng2024k2} and BB-GeoSFT/GeoPT~\cite{zhang2024bb}.

\subsection{Autonomous Driving}
Papers on autonomous driving use two main types of datasets: real-world visual datasets and simulation environments. Real-world datasets typically covering thousands of driving scenes for perception, planning, and decision-making tasks. 
Simulation platforms complement these by enabling configurable testing for specific scenarios and edge cases. 
Evaluation metrics in this field fall into three categories: perception, planning, and decision-making. Perception metrics include Mean Average Precision (MAP), Intersection-over-Union (IoU), and displacement errors (ADE/FDE)~\cite{li2022bevformer,hu2023planning}. Planning metrics assess trajectory accuracy (L2 Error), safety (Collision Rate)~\cite{tian2024drivevlm}, and overall performance (Route Completion, Driving Score)~\cite{mao2023language}. Language-based components are evaluated using automatic metrics (BLEU, ROUGE, CIDEr) and human evaluations for reasoning quality. Specialized metrics like Success Rate and Success Steps~\cite{wen2024dilu} measure system robustness in complex scenarios.

%% file: UCLLM_future.tex
\section{Future Prospects}
\label{sec:future}
LLMs have demonstrated significant value across urban computing applications, showing substantial potential in enhancing urban management. This includes improving the performance of urban indicator predictions and substituting for human experts in complex decision-making. Looking ahead, we anticipate that future research will focus on enhancing the generalization capabilities, interpretability, efficiency, and automatic planning of LLMs. Such advancements are crucial for elevating the intelligence of urban management systems.


\subsection{Enhance Generalization Capabilities}
Existing works typically expand the capabilities of LLMs through model fine-tuning. However, most studies are limited to small-scale datasets and models, which restricts their domain generalization capabilities. Given the excellent scalability of LLMs, fine-tuning models with larger parameter scales and increased data volumes can further enhance their domain generalization capabilities. Additionally, incorporating diverse training strategies, such as reinforcement learning from human feedback, introducing noise into prompts, and employing chain-of-thought generation, can help LLMs effectively utilize vast amounts of data and further enhance their performance.

\subsection{Improve Interpretability}
Improving the interpretability of LLMs in urban computing is essential for enhancing the decision-making processes in smart city management. Urban computing leverages various data sources, from traffic patterns to environmental sensors, requiring LLMs to not only predict but also explain their predictions for transparency and trustworthiness. By focusing on interpretability, developers can ensure that the models provide actionable insights that are understandable by city planners and the public alike. This improvement can lead to more effective urban planning strategies, better resource allocation, and increased public engagement by making the underlying decision processes accessible and justifiable. Moreover, adopting explainable AI frameworks helps in addressing potential biases and ensuring that the solutions proposed by LLMs adhere to ethical guidelines, ultimately leading to more sustainable and livable urban environments. 
Strategies include training models on annotated datasets with rationales for answers, integrating knowledge graphs to represent domain relationships, employing prompt engineering for step-by-step outputs, and leveraging human-in-the-loop systems to iteratively refine explanations for better transparency and usability.

\subsection{Increase Prediction Efficiency}
LLMs are highly beneficial for smart city management, yet their broader implementation is hindered by challenges related to computational efficiency. The extensive parameter counts of these models necessitate advanced computing hardware, leading to significant cost overheads. Furthermore, the sensitive nature of government and private data poses data security risks when these data are processed through cloud-based models. Consequently, enhancing the computational efficiency of these models while preserving their essential functionalities is a crucial area for future research.
Effective strategies to address these issues include model compression methods such as pruning and quantization, which reduce memory demands without compromising the models' performance on critical urban management tasks. Implementation of caching mechanisms for frequently accessed urban patterns and selective computation based on urban context importance further enhances prediction speed. Additionally, employing knowledge distillation allows the crucial insights of larger models (teacher models) to be instilled into smaller, more manageable models (student models). This adaptation facilitates the deployment of compact models at the edge of networks, addressing both efficiency and security concerns in urban computing environments.


%% file: UCLLM_conclusion.tex
\section{Conclusion}
\label{sec:conclusion}
The integration of Large Language Models (LLMs) into urban computing has driven significant advancements in data analysis and decision-making. In this review, we explored the diverse applications of LLMs, ranging from enhancing public transportation to optimizing autonomous driving in smart cities. We outlined key tasks, traditional approaches, and LLM implementation methodologies, highlighting their role in improving performance and driving innovative solutions to urban challenges. Building on these insights, we proposed viable solutions to address contemporary urban challenges. Furthermore, we organized existing domain-specific datasets and evaluation methods to assist in ongoing research efforts. Finally, the review explored the current limitations of LLMs in urban computing and outlined future research directions. LLMs in urban computing are still in their nascent stages, yet they hold immense potential to enable smarter, more connected cities.